\documentclass[pre,twocolumn,showpacs,preprintnumbers,amsmath,amssymb,longbibliography,floatfix]{revtex4}
\usepackage{amssymb,amsmath}
\usepackage{amsmath,amssymb}
\usepackage{float}
\usepackage{graphicx}
\usepackage{psfrag}
\usepackage{color}
\usepackage{dcolumn}
\usepackage{bm}
\usepackage[normalem]{ulem}

\def\beq{\begin{equation}}
\def\eeq{\end{equation}}
\def\bea{\begin{eqnarray}}
\def\eea{\end{eqnarray}}
\extrafloats{1000}
\begin{document}
\title { Dynamic scaling in the quenched disordered classical $N$-vector model}
\author{Sudip Mukherjee}\email{sudip.bat@gmail.com, sudip.mukherjee@saha.ac.in}
\affiliation{Barasat Government College,
10, KNC Road, Gupta Colony, Barasat, Kolkata 700124,
West Bengal, India}
\affiliation{Condensed Matter Physics Division, Saha Institute of
Nuclear Physics, Calcutta 700064, West Bengal, India}
\author{Abhik Basu}\email{abhik.123@gmail.com, abhik.basu@saha.ac.in}
\affiliation{Condensed Matter Physics Division, Saha Institute of
Nuclear Physics, Calcutta 700064, West Bengal, India}
\begin{abstract}
 We revisit the effects of short-ranged random quenched disorder on the universal scaling properties of the classical $N$-vector model with cubic anisotropy. We set up the nonconserved relaxational dynamics of the model, and study the universal dynamic scaling near the second order phase transition. We extract the critical exponents and the dynamic exponent in a one-loop dynamic renormalisation group calculation with short-ranged isotropic disorder. We show that the dynamics near a critical point is generically slower when the quenched disorder is relevant than when it is not, independent of whether the pure model is isotropic or cubic anisotropic. We demonstrate the surprising thresholdless instability of the associated universality class due to perturbations from rotational invariance breaking quenched disorder-order parameter coupling, indicating breakdown of dynamic scaling. We speculate that this may imply a novel first order transition in the model, induced by a symmetry-breaking disorder.

 \end{abstract}

\maketitle

\section{Introduction}
The large-scale, macroscopic effects of
disorder in statistical models and related condensed matter systems have
been active fields of theoretical research for a long time by now. In general, there are two kinds of 
disorders that can exist in a system, {\em viz.}, annealed and quenched
disorders, which are distinguished by their respective time-scales. In  systems with annealed disorders the impurities can
diffuse freely,  { eventually reaching the states with equilibrium distributions}. In contrast, for systems with a quenched
disorders the impurities are fixed in particular positions { and have no time-dependence}.  Consequently, the disorder { distribution cannot not reach}  thermodynamic equilibrium, { or does not ``thermalise''}. Past
studies on the effects of random quenched disorders on pure model systems,
e.g., the $O(N)$ model~\cite{mw,bc,wh} and self-avoiding walks on
random lattices~\cite{ck,kim1,rc,kim2} clearly illustrate the modifications in the universal critical behaviour due to the quenched impurities. The relevance of quenched
disorders on the universal scaling properties of the pure system,
is decided by  the well-known {\em Harris
criterion} \cite{harris}.  Perturbative renormalisation
group (RG) calculations on $O(N)$ symmetric models have shown that the
universal scaling properties have no dependence on the amplitude of the disorder
variance. For instance with short-ranged disorder, the scaling
exponents, even when they are affected by the disorder, are independent of the strength of the disorder variance~\cite{lub,replica,aharony}.

 In this article, we revisit the issue of the effects of the short-ranged random quenched disorder on the classical $N$-vector model with cubic anisotropy. In particular, we study the nonconserved relaxational dynamics of the model near its second order transition with generic couplings between the quenched disorder and the order parameter.
We concentrate on the {\em
random critical temperature} disorder which arises due to the
presence of a small amount of impurities or random bonding and which, { in the event of a second order transition}, 
can cause a variation in the critical transition temperature $T_c$
locally~\cite{lub,replica,aharony,krey}. We first construct a Landau mean-field theory to study the generic phase transitions in the model, { and look at the role of the symmetry-breaking disorder distribution in these transitions}. We then systematically set up a renormalised dynamic perturbation theory to study the time-dependent correlation functions near the critical point. We then calculate the dynamic renormalisation group (RG) fixed points with isotropic disorder - order parameter coupling, and extract the corresponding critical scaling exponents and the dynamic exponent within a one-loop perturbation theory, { or equivalently, at the linear order in $\epsilon\equiv 4-d$, where $d$ is the dimensionality of the space.  

{ The principal results from this work are as follows:} (i) We show that when the disorder is relevant (in a RG sense), i.e., with $N<4$, the dynamics near the second order transition is generically {\em slower} than that in the pure model: We find that the dynamic exponent $z=2+{\cal O}(\epsilon)$, making it larger than its value in the corresponding pure model where $z=2+{\cal O}(\epsilon^2)$~\cite{halpin} for small $\epsilon$. Thus disorder makes the dynamics slower near the critical point. { This holds whether or not the model in its pure limit is isotropic or cubic anisotropic.} The static critical exponents that we obtain from our perturbative dynamic RG agree with those obtained Refs.~\cite{lub,disorder-free}. (ii) We then  show that quenched disorder with rotational symmetry breaking couplings between the disorder and the order parameter are {\em relevant} perturbations on the RG fixed points,  with the RG flow trajectories running off to infinity, making the universality class of the model with isotropic disorder - order parameter coupling unobservable}. { Again this is generally true whether the corresponding pure model is isotropic or cubic anisotropic. This instability  indicates breakdown of dynamic scaling and is found to be {\em thresholdless}, i.e., any amplitude of the perturbation, however small, distabilises the RG fixed point for isotropic disorder with the RG flow lines running off to infinity, without the appearance of any new stable fixed point. (iii) Lastly, we construct a mean-field argument to speculate that this may imply a first order phase transition.}


The remainder of this article is organized as follows. In Sec.~\ref{model} we specify the relaxational equation of motion and the associated disorder distribution. Next in Sec.~\ref{mf}, we construct a Landau mean-field theory. Then in Sec.~\ref{scaling}, we set up a dynamic renormalisation group  analysis of the model to extract the universal scaling near the critical point.  We also obtain the dynamic exponent at the stable RG fixed points, which describes the dynamic scaling of the time-dependent correlation functions.  We show that a rotational symmetry breaking disorder order parameter coupling distabilises this stable fixed point for any nonzero amplitude of the symmetry breaking coupling. We then construct a mean-field argument in Sec.~\ref{first} that in the presence of a symmetry breaking coupling, In Sec.~\ref{summ} we discuss and summarise our results. We provide some technical details in Appendix for interested readers.


\section{Model} \label{model}

We start from the quenched disordered version of the well-known classical $N$-vector model with cubic anisotropy~\cite{aharony1,chaikin}. The Ginzburg-Landau free energy ${\cal F}=\int d^dx f$, where the free energy density $f$ is
given by~\cite{disorder-free}
\begin{equation}
 f=\sum_{i=1}^N\left[\frac{1}{  2}\{r_i({\bf x})\phi_i^2 +
(\nabla\phi_i)^2\} + u(\phi_i^2)^2 + v\phi_i^4\right], \label{free}
\end{equation}
where $\phi_i$ is the $N$-component ($i=1,...,N$) field.    Further, $u>0$ and
$v>0$ are the bare nonlinear coupling constants in the model. The cubic anisotropic terms
represent 
the breaking of the $O(N)$ symmetry by
the underlying crystal lattice, in particular, when in the ordered
phase the magnetisation prefers either one of the diagonals or the
edges of a hypercubic lattice~\cite{aharony1,chaikin}. In order to expand the scope of our study, we allow the quenched disorder  to couple with the order parameter field $\phi_i$ in a non-rotationally invariant way. For simplicity, we model this by considering the disorder as a sum of two parts - one part $\psi({\bf x})$ that couples with $\phi_i^2({\bf x})$ in a {\em rotationally invariant} way, where as the other part $\delta r_i({\bf x})$ depends on the index $i$ manifestly breaks the microscopic rotational invariance in the order parameter space, that is broken also by the $v\phi_i^4$-term in (\ref{free}). { We set $r_i({\bf x})=r_0 +\psi({\bf x}) +\delta r_i({\bf x})$, $r_0=(T-T_c)/T_c$ with $T$ and $T_c$ being the
temperature and the mean-field critical temperature, respectively.
Stochastic function $r_i({\bf x})$ represents the coupling of the field $\phi_i$ with the
disorder, such that $T_c^L=T_c-\psi(x)-\delta r_i({\bf x})$ is the local {\em
fluctuating } critical temperature for $\phi_i$.} For $v=0$ and for all $ \delta r_i({\bf x})= 0$, the
microscopic rotational invariance in the order parameter space is
restored and we get back the usual $O(N)$ model with quenched
disorder. On the other hand, if all $r_i=0$ and $v\neq 0$, it
reduces to the well-known pure cubic anisotropic model
\cite{aharony1,chaikin}.  Notice that  even when the pure limit of our model is isotropic ($v=0$), the disordered model breaks the rotational symmetry due to the presence of $\delta r_i$.

{
In order to completely define the model, we now specify the disorder distribution. We assume $\psi({\bf x})$ and $\delta r_i({\bf x})$ to be Gaussian-distributed with zero-mean and variances
\begin{eqnarray}
 \langle \psi({\bf x}) \psi(0)\rangle &=& 2D\delta^d({\bf x}),\nonumber \\
 \langle \delta r_i({\bf x}) \delta r_i(0)\rangle &=&2\hat D \delta^d({\bf x}).\label{disorder-dis}
\end{eqnarray}
Thus the quenched disorder is {\em short-ranged}. Disorder distributions (\ref{disorder-dis}) are a variant of what was used in Ref.~\cite{niladri}. 
If $\hat D=0$, the above distribution reduces to those used in \cite{lub,disorder-free};  see also Ref.~\cite{aharony} in this context.
The form of the variances in (\ref{disorder-dis}) ensures that both $D$ and $\hat D$ are equally {\em relevant} in a scaling sense. Clearly, that {\em both} nonzero $v$ and $\hat D> 0$ imply that the rotational invariance in the order parameter space is manifestly broken.  

\section{Mean-field theory}\label{mf}

It is instructive to first consider a Landau mean-field like description for this disordered system. We start from the free energy (\ref{free}). The corresponding partition function
\begin{equation}
 Z=\int {\cal D\phi}_i\exp [-\beta F].
\end{equation}
Systematic investigations of  the properties of systems
with quenched disorder requires averaging of the free energy functional
over the disorder distribution. This is conveniently done using
the replica method \cite{replica} resulting in an effective free
energy functional $\mathcal{F}_{eff}$, that consists of $M$ replicas
of the original order parameter fields. The average free energy is
then obtained from this $\mathcal{F}_{eff}$ in the limit of
$M\rightarrow 0$.
Then the thermodynamic free
energy averaged over the disorder distribution can be written as
\bea && F \equiv -\langle\ln Z\rangle = \lim_{m\rightarrow
0}\left[\frac{\langle Z^m\rangle-1}{m}\right]_{avg} \nonumber \\
&&= \lim_{m\rightarrow 0} \left\langle\left[
\frac{\prod_{\alpha=1}^M\prod_{i=1}^N {\cal D}{\{\phi_i^\alpha\}}
\exp[-\beta\mathcal{F}(\phi_i^\alpha)]-1}{m}\right]\right\rangle.\nonumber \\ \eea
 Here, angular brackets $\langle..\rangle$ represents averages over disorder
 distributions, $\alpha=1,2,....,M$ are the replica indices and
$\{\phi_i^\alpha\}$ represents $M$ replications of the order
parameters $\phi_i$; suffix $avg$ refers to averaging over the
disorder distribution. The effective free energy functional then reads
 \bea \exp[-\beta\mathcal{F}_{eff}]= \left\langle
\prod_\alpha{\cal D\phi}^\alpha_i \exp[-\beta\mathcal{F}
(\phi_i^\alpha)]\right\rangle_{avg}. \eea
 The corresponding $M$-replicated disorder averaged partition
 function $Z$ is written as
 \begin{widetext}
  
\begin{eqnarray}
 \langle Z^M\rangle &=& \prod_{\alpha=1}^M\prod_{i=1}^N\int {\cal D}\phi_{i\alpha} \exp \left[-\beta\int d^dx  f_{rep}+\beta\int d^dx f_{dis-rep}\right],\nonumber \\
 f_{rep} &=& \frac{1}{2}\sum_{i=1}^N\sum_{\alpha=1}^M\left[r_0\phi_{i\alpha}^2+({\boldsymbol\nabla}\phi_{i\alpha})^2\right]+u\sum_{\alpha=1}^M\left(\sum_{i=1}^N\phi_{i\alpha}^2\right)^2 + v \sum_{i=1}^N \sum_{\alpha=1}^M\phi_{i\alpha}^4,\nonumber \\
 f_{dis-rep}&=& \sum_{i,j=1}^N\sum_{\alpha,\beta=1}^M\phi_{i\alpha}({\bf x})^2D\phi_{j\beta}({\bf x})^2 
 + \sum_{i=1}^N\sum_{\alpha\beta=1}^M\phi_{i\alpha}({\bf x})^2\hat D\phi_{i\beta}({\bf x})^2. \label{rep-free1}
\end{eqnarray}

 \end{widetext}
 With $\hat D=0$, the above disorder-averaged free energy unsurprisingly reduces to that given in Ref.~\cite{disorder-free}.  With $\hat D>0$, (\ref{rep-free1}) naturally does not have any rotational invariance, but instead is cubic anisotropic~\cite{aharony}.
 
 In a mean-field description, $\phi_i({\bf x})$ is spatially constant. We assume below $T_c$ ($r_0<0$) $\phi_1=m\neq 0$ orders; all other $\phi_i,\,i=2,..,N$ vanish. In this mean-field approximation,
 \begin{eqnarray}
  f_{avg}&=&M\left[\frac{r_0}{2}m^2 + um^4 + vm^4\right],\\
  f_{dis-avg}&=&\frac{M(M+1)}{2}\left[Dm^4+\hat Dm^4\right].
 \end{eqnarray}
Now using the standard result $Lt_{M\rightarrow 0}\frac{Z^M-1}{M}=\ln Z$, we for the Landau mean field energy $f_{L}$
\begin{equation}
 f_L=\frac{r_0}{2}m^2+ \tilde u m^4,\label{landau}
\end{equation}
where, $\tilde u =u+v -(D+\hat D)/2$. Thus, $\tilde u$ can be positive, negative or zero. Following the standard argument~\cite{chaikin}, we find that (\ref{landau}) implies a second order transition at $T=T_c$, that is identical to the pure mean-field Ising universality class for $\tilde u>0$. This holds  for small $D$ and $\hat D$. If $\tilde u<0$, $f_L$ is unbounded from below and hence thermodynamically unstable. To stabilise, as is usual, we add a $u_6 m^6$ term in (\ref{landau})~\cite{chaikin}: $f_L=\frac{r_0}{2}m^2- |\tilde u| m^4 + u_6 m^6$. This gives
a first order transition at $T^*=T_c+\frac{\tilde u^2}{16 u_6}$~\cite{chaikin} and a tricritical point at $\tilde u=0$. { Thus, within this simple mean-field theory, by tuning $\hat D$ and with $u,\,v,\,D$ fixed, i.e., by tuning the degree of rotational symmetry breaking in the disorder distribution, the phase transition can be changed from a second order to a first order transition through a tricritical point.} Of course, these predictions may not hold, as we know that in general fluctuations could be important below the upper critical dimension, changing the universal properties near the critical point. Fluctuations can also change a mean-field second order transition to a first order one~\cite{coleman}. To study this, we need to systematically account for the fluctuations that we set out to do below.

\section{Universal dynamical scaling near second order transition}\label{scaling}

We note that for sufficiently small $D,\,\hat D$, the mean-field theory above predicts a second order transition at $r_0=0$. 
We are interested in the time-dependent statistical mechanics of the spins $\phi ({\bf x},t)$, for which we must start from an appropriate dynamical equation of motion. For simplicity, we focus on the non-conserved relaxational dynamics of the spins $\phi({\bf x},t)$, and in the absence of any other hydrodynamic degree of freedom, we construct the Model A (in the language of Ref.~\cite{halpin}; { see Ref.~\cite{folk} for more recent review on critical dynamics})  equation of motion for $\phi_i({\bf x},t)$:
\begin{eqnarray}
 \frac{\partial \phi_i}{\partial t} = -\Gamma \frac{\delta {\cal F}}{\delta\phi_i} + \eta_i, \label{modelA}
\end{eqnarray}
where $\Gamma>0$ is the kinetic coefficient and $\eta_i$ is a thermal noise that is assumed to be Gaussian-distributed with zero mean. Since the system is in equilibrium, the variance of $\eta_i$ can be fixed by using the Fluctuation-Dissipation Theorem (FDT)~\cite{chaikin}. We find
\begin{equation}
 \langle \eta_i({\bf x},t)\eta_j(0,0)\rangle = 2\Gamma T \delta_{ij}\delta^d({\bf x})\delta(t), \label{thermal-noise}
\end{equation}
where we have set the Boltzmann constant $k_B=1$. By using (\ref{free}), the explicit form of (\ref{modelA}) becomes
\begin{equation}
 \frac{1}{\Gamma}\frac{\partial\phi_i}{\partial t} = - [r_i ({\bf x})\phi_i - \nabla^2 \phi_i +  4u \sum_{j}\phi_i \phi_j^2 + 4v\phi_i^3] + \frac{\eta_i}{\Gamma},\label{eom1} 
\end{equation}
{ $i,j=1,..,N$. Equation~(\ref{eom1}) together with the disorder distribution variance~(\ref{disorder-dis}) fully specify the dynamical model.}

\subsection{Second order transition: Universal critical exponents} \label{scaling1}

Classical cubic anisotropic $N$-vector model with free energy (\ref{free}) without any quenched disorder undergo a second order phase transition at the critical temperature, which in a mean-field description is at $T=T_c$ or $r_0=0$. The model exhibits universal scaling for the thermodynamic quantities and the equal-time correlation function of the spin fluctuations at the vicinity of the critical point. The critical exponents have been evaluated within perturbative dynamic RG frameworks~\cite{chaikin}. Further, near the critical point the time-dependent spin response function and spin correlation function display universal dynamical scaling as well; see, e.g., Ref.~\cite{halpin,folk}. Short-ranged quenched disorder is known to affect the universal scaling of the thermodynamic quantities and equal-time correlation functions~\cite{kim1,lub}, consistent with the Harris criteria~\cite{harris} for the present model.

We take the dynamical route to study the effects of short-ranged quenched disorder on the dynamic universality of the classical $N$-vector model near its critical point. This allows us to find whether or not quenched disorder affects the dynamic scaling exponent of the pure (i.e., without any quenched disorder) model. { It may be noted that this dynamic approach no longer necessitates introduction of the replica method to calculate the universal critical exponents. In addition, this method directly gives the dynamic exponent $z$.}

\subsection{Dynamic RG analysis}

It is convenient to express (\ref{eom1}) as the generating functional of the correlation functions and then average over the disorder distribution. The resulting disorder averaged generating functional has the form
\begin{equation}
 {\cal Z}=\int {\cal D}\phi_i {\cal D}\hat\phi_i \exp ({\cal - S}),
\end{equation}
where the action functional ${\cal S}$ reads
\begin{widetext}

\begin{eqnarray}
 {\cal S}&=&\int d^dx dt\left[(\hat\phi ({\bf x},t))^2 \frac{T}{\Gamma} + \hat\phi_i \{\frac{\partial_t \phi_i}{\Gamma} +r_0\phi_i-\nabla^2\phi_i + 4v\phi_i^3 + 4u\phi_i\phi_m^2\}\right] \nonumber \\ &-&\int d^dx dt_1 dt_2 \left[\phi_i({\bf x},t_1)\hat\phi_i({\bf x},t_1)D \phi_j({\bf x},t_2)\hat\phi_j({\bf x},t_2) + \phi_i({\bf x},t_1)\hat\phi_i({\bf x},t_1) \hat D \phi_i({\bf x},t_2)\hat\phi_i({\bf x},t_2)\right],\label{action}
\end{eqnarray}

\end{widetext}
with $i,j,m=1,..,N$. 
Here, $\hat\phi_i({\bf x},t)$ is the dynamic conjugate field of $\phi_i({\bf x},t)$~\cite{janssen}. 
Clearly, the disorder term $\hat D$ violates the rotational invariance of $\cal S$, so does the $v$-term in (\ref{action}). The presence of the anharmonic terms (i.e., the $u$-, $v$-, $D$-, $\hat D$-terms in (\ref{action})) precludes any exact enumeration of the correlation function. This necessitates a perturbative approach. It turns out that the na\"ive perturbative theory actually produces diverging corrections to the model 
parameters in (\ref{action}). This calls for a systematic dynamic RG analysis.

The momentum shell dynamic RG procedure { is well-documented in the literature~\cite{halpin,uwe-book}; { see also Ref.~\cite{vasilev} for a detailed exposition of the applications of the dynamic RG techniques for critical phenomena and critical dynamics, including effects of quenched disorders}. We nevertheless give below a brief outline of it for the convenience of the reader.} It consists of tracing over the short wavelength 
Fourier modes
of $\phi_i({\bf x},t)$ and $\hat\phi_i({\bf x},t)$, followed by a rescaling of lengths and time.
More precisely, we  follow the usual approach of initially restricting wavevectors  
to be confined in a $d$-dimensional Brillouin zone: $|{\bf q}|<\Lambda$, where 
$\Lambda$ is an
ultra-violet cutoff,  which should be of order the inverse of the lattice spacing $a$. The  precise value of $a$ should have no effect on the universal scaling of the model. The fields ${\hat
\phi_i}({\bf x},t)$ and $\phi_i({\bf x},t)$
are separated into high and low wave vector parts
$\phi_i({\bf x},t)=\phi_i^>({\bf x},t)+\phi_i^<({\bf x},t)$ and $\hat\phi_i({\bf x},t)=\hat\phi_i^>({\bf x},t) + \hat\phi_i^<({\bf x},t)$,
where $\phi_i^>({\bf x},t)$ and $\hat\phi_i^>({\bf x},t)$ have support in the large wave vector  (short wavelength) range $\Lambda
e^{-\delta l}<| {\bf q}|<\Lambda$, while $\phi_i^<({\bf x},t)$ and $\hat\phi_i^<({\bf x},t)$ have support in the small 
wave vector (long wavelength) range $|{ \bf q}|<e^{-\delta l}\Lambda$; { here, $\delta l\ll 1,\,b\equiv\exp(\delta l)\approx 1+\delta l$}.
We then integrate out $\phi_i^>({\bf x},t)$ and $\hat\phi_i^>({\bf x},t)$. This integration is done perturbatively 
in  the anharmonic couplings in (\ref{action}). It is well-known that  this perturbative expansion 
can be represented by Feynmann diagrams~\cite{halpin,uwe-book,vasilev}, with the order of perturbation theory 
reflected by the number of loops in the diagrams we consider. After this 
perturbative step, we rescale lengths
with ${\bf x}={\bf x }' e^{\delta l}$,  in order to restore the UV cutoff back to 
$\Lambda$,  and also time with $t=t'e^{z\delta l}$.  This procedure is then followed
by usual rescaling of the long wave length parts of the fields $\phi({\bf x},t)$ and $\hat\phi({\bf x},t)$; see Appendix.   The Feynman graphs (or “vertices”) representing the anharmonic couplings
$4u \hat \phi_i\phi_i \phi_j^2,\,4v\hat\phi_i \phi_i^3,\,\hat\phi_i\phi_i D\hat \phi_j\phi_j,\,\hat\phi_i \phi_i\hat D\hat\phi_i\hat\phi_i$ are shown in Fig.~\ref{vertex} in Appendix.

The bare propagator and the correlation function in the Fourier space are
\begin{eqnarray}
 \langle \hat \phi_i(-{\bf k},-\omega)\phi_j({\bf k},\omega)\rangle &=& \frac{\delta_{ij}} {\frac{-i\omega}{\Gamma} + r_0 + k^2},\\
 \langle \phi_i({\bf k},\omega)\phi_j(-{\bf k},-\omega)\rangle &=& \frac{\delta_{ij}\frac{2T}{\Gamma}} {\frac{\omega^2}{\Gamma^2} + (r_0 + k^2)^2},
\end{eqnarray}
where $\bf k$ is a Fourier wavevector and $\omega$ is a frequency.

We restrict ourselves here to a one-loop approximation. At this order the propagator receives four fluctuation corrections, originating from nonzero $u,\,v,\,D$ and $\hat D$ respectively; the relevant Feynman diagrams are given in Fig.~\ref{prop} in Appendix.

\begin{widetext}

Evaluation of the Feynman diagrams in Fig.~(\ref{prop}) in Appendix allows us to extract the following fluctuation corrected model parameters:

\begin{eqnarray}
 r_0^<&=& r_0 - \left[ u\{12 + 4(N-1)\} +12v\right]\int \frac{d^dq}{(2\pi)^d}\frac{1}{r_0+q^2} +2\left(D+\hat D\right) \int \frac{d^dq}{(2\pi)^d}\frac{1}{r_0+q^2},\label{req}\\
 \left(\frac{1}{\Gamma}\right)^< &=& \frac{1}{\Gamma} +2\frac{D+\hat D}{\Gamma} \int \frac{d^dq}{(2\pi)^d}\frac{1}{(r_0+q^2)^2}. \label{gammaeq}
\end{eqnarray}
Notice that there are non-vanishing disorder-induced fluctuation corrections to $\Gamma$ already at the one-loop, unlike in the pure model, where such corrections appear only at the two-loop order (we do not show that here).

Likewise, $u,\,v,\,D$ and $\hat D$ are each renormalised at the one-loop order by the Feynman diagrams as shown in Fig.~\ref{u-diag}, Fig.~\ref{v-diag}, Fig.~\ref{D-diag} and Fig.~\ref{hat-D-diag}, respectively, in Appendix. The calculations of the Feynman graphs in Fig.~\ref{prop}, Fig.~\ref{u-diag}, Fig.~\ref{v-diag} and Fig.~\ref{D-diag} and Fig.~\ref{hat-D-diag} are given in Appendix. The resulting fluctuation-corrected parameters  $u^<,\,v^<,\,D^<$ and $\hat D^<$ are given in Eqs.~(\ref{ueq}), (\ref{veq}), (\ref{Deq}) and (\ref{hatDeq}), respectively, below. 


 
 \begin{eqnarray}
  u^< &=& u-u^2\left[36+4(N-1)\right]\int_{\Lambda/b}^\Lambda\frac{d^dq}{(2\pi)^d}\frac{1}{(r_0 + q^2)^2} -24uv \int_{\Lambda/b}^\Lambda\frac{d^dq}{(2\pi)^d}\frac{1}{(r_0 + q^2)^2} \nonumber \\&+&u \left(12D+4\hat D\right) \int_{\Lambda/b}^\Lambda\frac{d^dq}{(2\pi)^d}\frac{1}{(r_0 + q^2)^2},\label{ueq}
 \end{eqnarray}





%
\begin{eqnarray}
 v^<&=&v-\left[48uv +36v^2\right] \int_{\Lambda/b}^\Lambda\frac{d^dq}{(2\pi)^d}\frac{1}{(r_0 + q^2)^2} + 12 v \left(D+\hat D\right) \int_{\Lambda/b}^\Lambda\frac{d^dq}{(2\pi)^d}\frac{1}{(r_0 + q^2)^2},\label{veq}
\end{eqnarray}




 
\begin{eqnarray}
 D^<&=&D- D\left[8(N+2)u +24v\right]\int_{\Lambda/b}^\Lambda\frac{d^dq}{(2\pi)^d}\frac{1}{(r_0 + q^2)^2} - 8u\hat D\int_{\Lambda/b}^\Lambda\frac{d^dq}{(2\pi)^d}\frac{1}{(r_0 + q^2)^2} \nonumber \\&+& \left(8D +4\hat D\right)D \int_{\Lambda/b}^\Lambda\frac{d^dq}{(2\pi)^d}\frac{1}{(r_0 + q^2)^2}, \label{Deq}
\end{eqnarray}

 




\begin{eqnarray}
\hat D^< = \hat D - \left(16 u + 24 v\right)\hat D \int_{\Lambda/b}^\Lambda\frac{d^dq}{(2\pi)^d}\frac{1}{(r_0 + q^2)^2} + \left(12 D + 8\hat D\right)\hat D \int_{\Lambda/b}^\Lambda\frac{d^dq}{(2\pi)^d}\frac{1}{(r_0 + q^2)^2}. \label{hatDeq}
\end{eqnarray}

\end{widetext}
After rescaling space, time and the fields $\phi_i$ (see Appendix), Eqs.~(\ref{req}-\ref{hatDeq}) result into the following RG recursion relations

\vspace{10cm}

\begin{eqnarray}
 \frac{dr_0}{dl}&=&2r_0+4(N+2)u +12 v -2(D+\hat D),
 \end{eqnarray}
 \begin{equation}
 \frac{d\Gamma}{dl}=\Gamma\left[z-2- 2(D+\hat D)\right],\label{gamm-flow}
 \end{equation}
 \begin{equation}
 \frac{du}{dl}=u\left[\epsilon - 4(N+8)u- 24 v + 12 D + 4 \hat D\right],
 \end{equation}
 \begin{equation}
 \frac{dv}{dl}=v\left[\epsilon - 48u -36v +12 D + 12\hat D\right],\label{vlow}
 \end{equation}
 \begin{eqnarray}
 \frac{dD}{dl}&=&D\left[\epsilon - 8(N+2) u - 24 v+ 8D+4\hat D\right]\nonumber \\ &-& 8u\hat D,
 \end{eqnarray}
 \begin{eqnarray}
 \frac{d\hat D}{dl}&=&\hat D\left[\epsilon - 16u -24 v+12D+8\hat D\right] \label{hatDflow}.
\end{eqnarray}

Here, $\epsilon \equiv 4-d$. At the RG fixed point, $du/dl=0=dv/dl=dD/dl=d\hat D/dl$. Notice that with $\hat D=0$, the above RG recursion relations unsurprisingly reduce to those reported in \cite{lub,disorder-free}.  First we revisit the fixed points with $\hat D =0$ and discuss the associated dynamic scaling. The RG fixed points are well-known~\cite{lub,disorder-free}, that we reproduce for the sake of completeness and for the convenience of the reader. { We here focus on the following three stable fixed points given by $u^*,\,v^*,\,D^*$~\cite{lub,replica,disorder-free}:
\begin{itemize}
 \item Cubic anisotropic pure fixed point: $u^*=\epsilon/(12N),\,v^*=(N-4)\epsilon/(36N),\,D=0$ to the linear order in $\epsilon$. Disorder is irrelevant. Unsurprisingly, at this fixed point $z=2+{\cal O}(\epsilon)^2$. 
 \item Isotropic random fixed point: $u^*=\epsilon/[16(N-1)],\,D^*=(4-N)\epsilon/[16(N-1)],\,v=0$. Since $D^*$ should be positive, this fixed point exists for $N<4$. At this fixed point, $z=2+2(4-N)\epsilon/[16(N-1)]$ at this linear order in $\epsilon$~\cite{krey}, and larger than 2, its value in the noninteracting (harmonic) theory. In contrast, $z=2+{\cal O}(\epsilon)^2$ for $N\geq 4$. Since in the pure model, $z=2+ {\cal O}(\epsilon)^2$, the dynamic exponent when the disorder is relevant is larger than both when it is not for sufficiently small $\epsilon$, and the value of $z$ in the noninteracting (Gaussian) theory. Thus the corresponding relaxational dynamics is slower.
 \item Cubic anisotropic random fixed point: $u^*=\frac{\epsilon}{24(N-2)},\,v^*=(N-4)u^*/4,\,D^*=(4-N)u^*$. Since $D^*$ should be positive, this fixed point exists for $N<4$. At this fixed point, dynamic exponent $z=2+ 2(4-N)\epsilon/[24(N-2)]>2$ for $N<4$. On the other hand, for $N\geq 4$, $D^*=0$, and hence $z=2+{\cal O}(\epsilon)^2$. Again we conclude that the dynamic exponent when the disorder is relevant is larger than both when it is not for sufficiently small $\epsilon$, and the value of $z$ in the noninteracting (Gaussian) theory. 
\end{itemize}
The correlation function exponent $\eta$ remains zero at this order at all the three RG fixed points discussed above. Overall, thus disorder is irrelevant for $N\geq 4$, in agreement with the results of Refs.~\cite{lub,disorder-free}.}

We now study the case with $\hat D>0$.  Let us examine the stability of the nontrivial fixed point. For that we define  dimensionless numbers $\mu_1=\hat D/D$, $\mu_2=\hat D/u$ and also $\mu_3=\hat D/v$ (if the pure model is cubic anisotropic). Then
\begin{eqnarray}
 \frac{d\mu_1}{dl}&=&\frac{1}{D}\frac{d\hat D}{dl}-\frac{\hat D}{D^2}\frac{dD}{dl}\nonumber\\
 &=& \mu_1[\epsilon - 16u -24 v + 12 D + 8\hat D] \nonumber \\&-& \mu_1 [\epsilon - 8(N+2) u - 24 v - 8u \mu_1 + 8D + 4\hat D]\nonumber \\
 &=&\mu_1 [ 4D +4\hat D + 8N u + 8 u\mu_1]>0,\label{mu1}\\
 \frac{d\mu_2}{dl}&=&\frac{1}{v}\frac{d\hat D}{dl}-\frac{\hat D}{v^2}\frac{d u}{dl}\nonumber \\
 &=&\mu_2 [\epsilon -16 u-24v +12D + 8\hat D\nonumber \\&+&(32 + 4N)u +24 v-12 D-4\hat D]\nonumber \\&=&
 \mu_2 [(16 + 4N)u + 4\hat D]>0.\label{mu2}
\end{eqnarray}
Thus, $\mu_1 =0$ and $\mu_2=0$ are the only fixed points of (\ref{mu1}) and (\ref{mu2}) respectively, both of which are, however,  unstable. Therefore, unless $\mu_1$ and $\mu_2$ are {\em exactly} zero (i.e., $\hat D=0$, corresponding to a perfectly isotropic disorder distribution), $\hat D$ continues to grow under coarse-graining and the associated RG trajectories eventually run off to infinity in the thermodynamic limit. { In Fig.~\ref{flow}, we show schematic plots of the RG trajectories flowing off to infinity for finite $\hat D$ in the $u-\hat D$ and $D-\hat D$ (for $N<4$) planes.
\begin{figure}[htb]
\includegraphics[width=8cm]{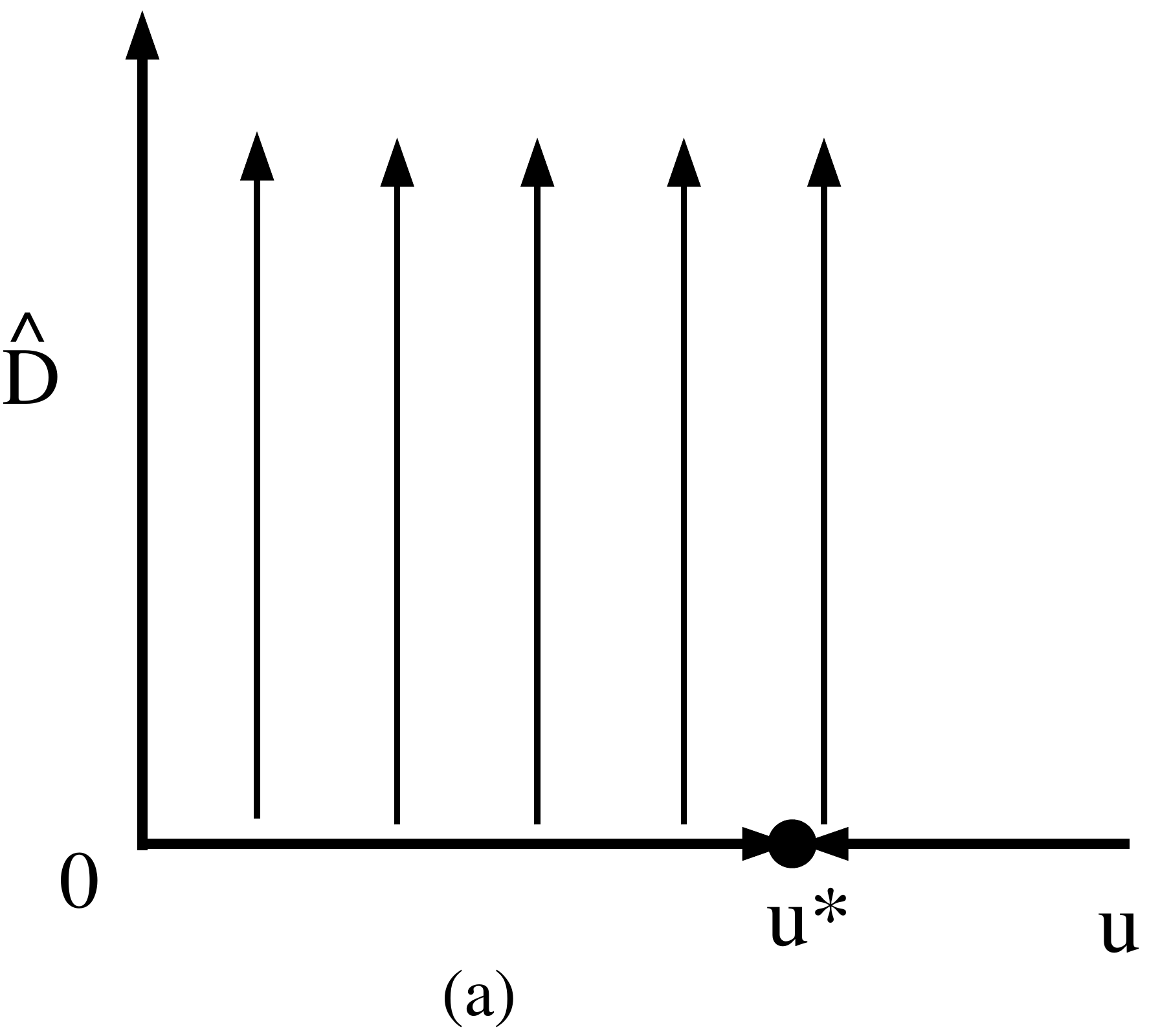}\\
\includegraphics[width=8cm]{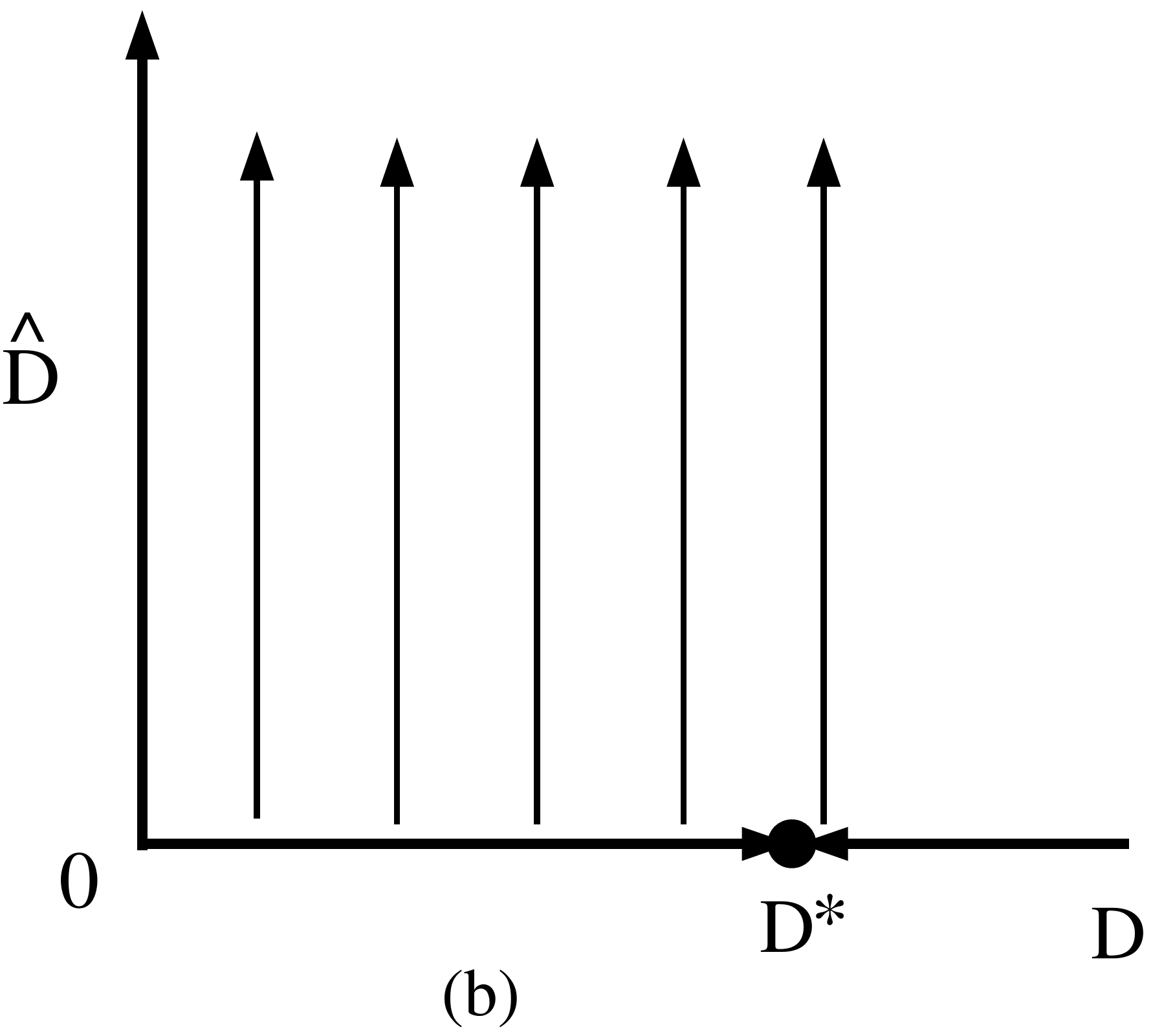}
\caption{Schematic RG flow diagrams in the (a) $u-\hat D$ plane, and (b) $D-\hat D$ plane. In each of these diagrams, RG trajectories starting  with any non-zero $\hat D$ (vertical lines with arrows) run off to infinity (see text). Symbols $u^*$ and $D^*$ represent the fixed point values of $u$ and $D$, respectively, for $\hat D=0$; these fixed points are stable (attractive) with $N<4$ (see text).  That the RG flow lines starting with finite $\hat D$ are shown to be parallel to each other is only for easier representation.}\label{flow}
\end{figure}
From (\ref{gamm-flow}), a diverging $\mu$ (or $\hat D$) implies $z\rightarrow \infty$, indicating breakdown of dynamic scaling.} 

{ Notice that when $v\neq 0$, the flow of $\mu_3$ reveals more interesting behaviour. We find
\begin{eqnarray}
 \frac{d\mu_3}{dl}&=&\frac{1}{v}\frac{d\hat D}{dl}-\frac{\hat D}{v^2}\frac{dv}{dl}\nonumber \\ 
 &=& \mu_3\left[\epsilon -16 u-24 v+12 D+8\hat D\right]\nonumber \\
 &-&\mu_3\left[\epsilon -48 u-36v +12 D +12 \hat D\right]\nonumber \\
 &=&\mu_3 \left[36 u + 12 v - 4\hat D\right].
\end{eqnarray}
Thus at the RG fixed point 
\begin{eqnarray}
 \hat D &=& 9u + 3v,\\
 \mu_3&\equiv& \frac{\hat D}{v}=3
\end{eqnarray}
at this stable RG fixed point:
Using the fact that $\mu_1\equiv \hat D/u$ diverges, i.e., $\hat D\gg u$, we get $\hat D = 3v$ at the RG fixed point. Notice that we can arrive at the above condition directly by using either (\ref{vflow}) or (\ref{hatDflow}) above and setting $u=0=D$ at the RG fixed point. In other words, if $u=0=D$ then any $\hat D>0$ and $v>0$ satisfying $\hat D=3v$ are fixed point values, or equivalently, $\mu_3=\hat D/v$ appears as a fixed ratio in the problem, i.e., the precise bare (unrenormalised) or microscopic values of $\hat D$ and $v$ may affect the large scale scaling behaviour. In fact, we apparently find from (\ref{gamm-flow}) that the dynamic exponent $z$ may now vary continuously with $\hat D$, a scenario usually not observed in equilibrium systems. How these results may change beyond the  one-loop calculation employed here remains to be seen. A schematic plot of the RG flows in the $v-\hat D$ plane is shown in Fig.~\ref{vlow} below.
\begin{figure}[htb]
\includegraphics[width=8cm]{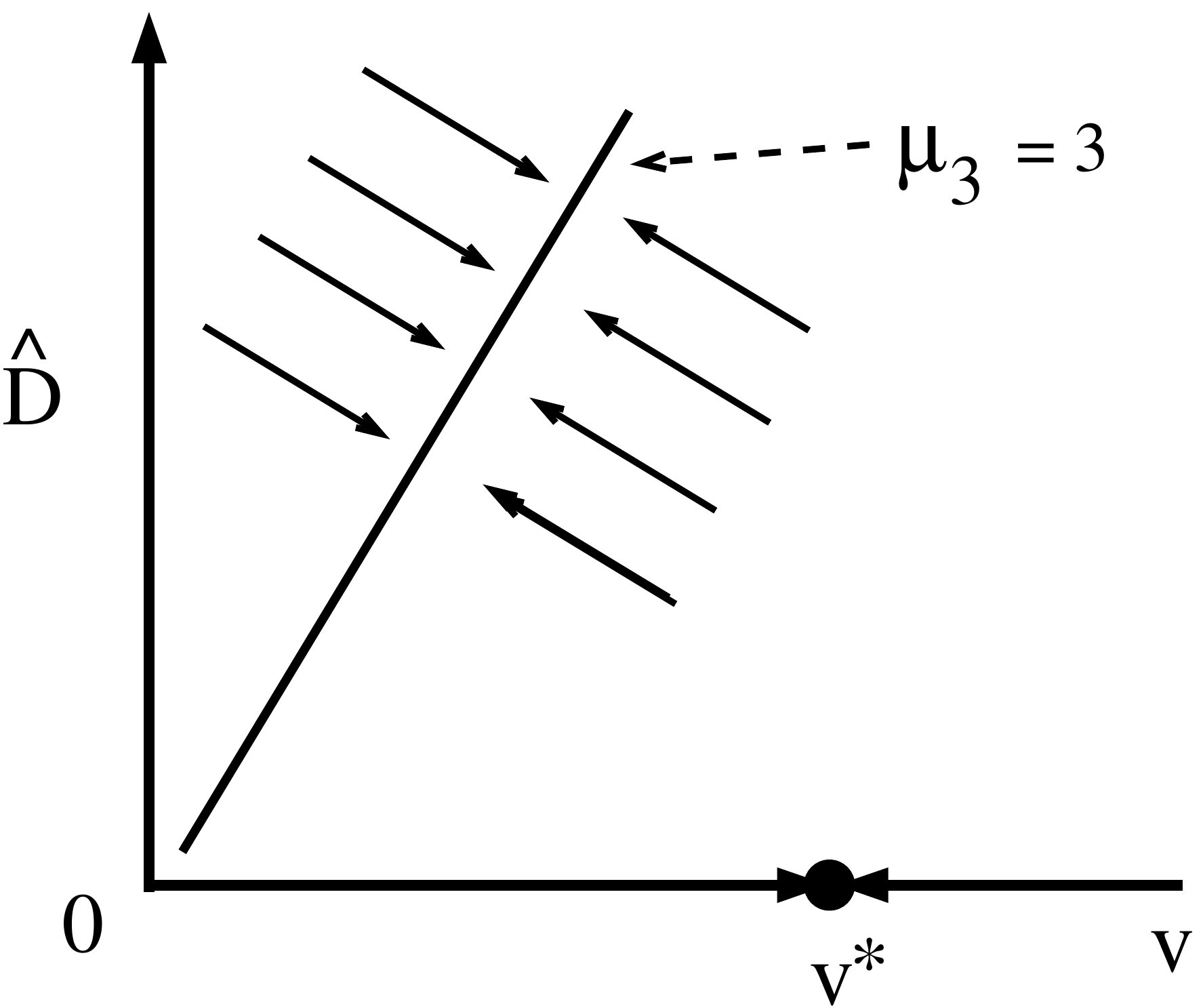}
\caption{Schematic flow diagram in the $v-\hat D$ plane. Here, $v^*$ is the RG fixed point value of $v$ with $u=0=D=\hat D$. The stable fixed line $\hat D=3v$, or $\mu_3=3$ is shown (see text).}\label{vflow}
 \end{figure}

}

\subsection{A first order transition?}\label{first}

{ Let us ask when $\hat D>0$, what does the RG flow lines running off to infinity  imply?}  { Such runaway RG flow trajectories indicate breakdown of dynamic scaling that is a hallmark of second order transitions.} We here heuristically speculate that this may actually indicate a first order transition. We analyse this by using the effective disorder-averaged Landau mean-field theory for the system constructed in Sec.~\ref{mf}. We note that for a very large $\hat D=3v$ and with both of $u$ and $D\sim {\cal O}(1)$, $\tilde u<0$ definitely. 
This then is known to describe a first order transition for $\tilde u <0$ and a tricritical point at $\tilde u=0$~\cite{chaikin}. At this simple mean-field level treatment, the scaling properties at the tricritical point of this disordered system is indistinguishable from  { a pure system with tricritical point with the same structure but without any disorder}. Fluctuations are expected to modify the universal scaling with possibly an imprint of the disorder in the form of a new universality class. The { basic} question that remains is whether $\hat D$ really becomes infinity, as predicted by the flow equations above, which would make the effective Landau free energy unbounded from below even in the presence of a $u_6 m^6$ term with a finite $u_6$. This is clearly unphysical. This is however an unlikely scenario given that near a first order transition, fluctuations are bounded and moreover there should be other higher order terms in the order parameter which are likely to saturate any growing $\hat D>0$. { On the other hand, we could alternatively interpret diverging $\mu_1$ and $\mu_2$ as $u\rightarrow 0$ and $D\rightarrow 0$ along with finite $\hat D=3v$ in the long wavelength limit. In that case, $\tilde u<0$ and a finite $u_6$ suffice to stop the effective Landau free energy getting unbounded from below. With $\tilde u<0$, a first order transition then automatically follows. Now if indeed we can neglect $u,\,D$ in favour of $v,\,\hat D$, we should end up having $N$ identical decoupled quenched disordered Ising models in the long wavelength limit, each of which undergoes a first order transition. Further investigations should be done to investigate this prediction. } In this context, 
it will be interesting to explore any connections between our speculation of a first order transition here and the prediction of the possibility of first order transitions in the two-dimensional classical XY model in a random symmetry breaking field~\cite{cardy1}. We do not discuss this here any further. We end this section with a note of caution. { While } it is { often} suggested that such an absence of a stable RG fixed point essentially implies a first order transition, although some studies~\cite{chandanda,john1,john2} have demonstrated that this lack of a stable RG fixed point is a problem of perturbative RG, valid strictly for small (bare) coupling constants. Applying duality transformations, Refs.~\cite{chandanda,john1,john2} did eventually found stable RG fixed points corresponding to second order phase transitions in the respective models. Whether and how such a duality transformation can be applied to the present model, and whether the dual model will have stable RG fixed points restoring second order transitions remain open questions, that we hope will be addressed in the future.  We would like to add here that our speculation for a first order transition is based only on mean-field like arguments. Alternatively, it could also be a smeared transition; see discussions in Ref.~\cite{lub}.

\section{Summary and outlook}\label{summ}

We have studied the  nonconserved  relaxational dynamics of the classical $N$-vector model with cubic anisotropy in the presence of quenched disorder. We focus on the short range random critical temperature, and allow for rotational symmetry breaking disorder distribution. We set up a dynamic renormalised perturbation theory. We first extract the critical scaling exponents and the dynamic scaling exponent near the critical point without any symmetry breaking disorder order parameter coupling within a one-loop  dynamic RG calculation. We show that when the disorder is relevant (i.e., for $N<4$) the dynamic exponent $z$ is larger than two, i.e., the quenched disorder slows down the relaxation vis-a-vis the noninteracting theory; $z$ is also found to be larger than its value in the pure model. This implies a generic disorder-induced slowing of the relaxational dynamics near the critical point. This holds independent of whether the pure model is isotropic or cubic anisotropic.  Notice that if we had imposed conserved relaxational dynamics, we would have found $z=4-\eta$~\cite{halpin,krey}. Since $\eta =0$ at the linear order in $\epsilon$, $z=4$ at the lowest order, for conserved relaxational dynamics, unchanged from its noninteracting limit value or its value in the associated pure model. Thus, quenched disorder affects the dynamic exponent {\em differently} in different dynamics, a result directly testable in relevant experiments or numerical simulations of appropriate lattice-based spins models.  Unsurprisingly, the static critical exponents obtained from our calculations match with the known results obtained from the standard static perturbation theory. We have thus established a direct one-to-one correspondence between the dynamic perturbation theory  { that we have constructed} and the known static perturbation theory. { We further show that the a non-zero amplitude $\hat D$ of a symmetry-breaking disorder order parameter coupling without any threshold makes the stable RG fixed point that controls the critical scaling with isotropic disorder as mentioned above unstable, possibly leading to breakdown of dynamic scaling. If the pure model is isotropic (i.e., $v=0$), then we show that a non-zero $\hat D$ leads to the RG flow trajectories running off to infinity without the appearance of any new stable RG fixed point.} { On the other hand, if the corresponding pure model is cubic anisotropic ($v\neq 0$), then $\hat D=3v$ appears as a RG fixed line, with $\hat D/u,\,\hat D/D$ diverging in the long wavelength limit. } We then construct a disordered averaged Landau mean-field theory to heuristically argue that such an instability may imply a first order transition. Thus, a novel symmetry-breaking disorder induced first order transition, that is otherwise second order in nature,  is an intriguing possibility in this model. The instability associated with the amplitude of the symmetry-breaking disorder is found to be {\em thresholdless}. 
Whether or not there is indeed a first order transition for $\hat D>0$ and if so, whether it is thresholdless, can be investigated by Monte-Carlo simulations of suitably-constructed lattice models with appropriate quenched disorders systematically. On the analytical front, whether applications of the duality transformation provide any insight to the phase transitions remain important open issues.

The present study can be extended in several ways. For instance, the role of  an additional hydrodynamic degree of freedom, non-critical or critical (Model C and Model D, respectively, in the nomenclature of Ref.~\cite{halpin}), and its interplay with a symmetry-breaking disorder could  be investigated. We have confined ourselves here with short-ranged quenched disorder. It would also be interesting to explore how long range quenched disorder may affect the results from this study~\cite{wh}. It would be interesting to see how nonperturbative RG, applied in the related problem of dilute Ising model~\cite{non-pert-RG} and conformal field theory methods~\cite{conformal}, apply to the current model.  For the $N=3$ case, i.e., for the classical Heisenberg model, the conserved dynamics of the spin model includes mode coupling terms~\cite{halpin,chaikin}, which affects the dynamic scaling near the critical point, but not the static critical exponents. Whether or how quenched disorder can couple with the mode coupling term in a ``relevant'' way (in a RG sense) should be explored.
The present study has been confined to thermal equilibrium, for which the structure of the dynamical equation is restricted so as to maintain the conditions of  { detailed balance}. 
It would be interesting to consider the issues studied here in the context of nonequilibrium or active systems. Nonequilibrium systems are known to be more prone to generic symmetry-breaking perturbations, opening up the possibilities of richer and more complex phase transitions and phase behaviour. { For instance, how random quenched disorder can affect the scaling of aging~\cite{raja} and steady states of an active or nonequilibrium $O(N)$ model should be an interesting future question to study.}

\section{Acknowledgement}

We thank J. Toner for discussions, A. Aharony for his comments on a preliminary version of this article, and N. Sarkar for providing technical support.

\appendix
\section{Derivation of the action functional}

We start from the equation of motion (\ref{eom1}). This may be cast as a generating functional
\begin{equation}
 {\mathcal Z}_{dis}=\int {\cal D}\phi_i\,{\cal D}\hat\phi_i\exp[-S_{dis}],
\end{equation}
where 
\begin{widetext}
 
\begin{equation}
 S_{dis}=-\int d^dx dt [\hat\phi_i({\bf x},t)]^2\frac{T}{\Gamma} +\int d^dx dt \hat\phi_i [\frac{1}{\Gamma} \partial_t\phi_i + (r_0 +\psi +\delta r_i)\phi_i -\nabla^2 \phi_i + 4u \phi_i \phi_j^2 + 4v \phi_i^3],
\end{equation}
with $i,j=1,..,N$.
\end{widetext}
We now average ${\mathcal Z}_{dis}$ over the Gaussian distributed $\psi$ and $\delta r_i$ having variances (\ref{disorder-dis}). Noting that the quenched disorder fields $\psi({\bf x})$ and $\delta r_i ({\bf x})$ are time independent, and so are the variances (\ref{disorder-dis}), averaging over the Gaussian distributions of $\psi$ and $\delta r_i$ directly yield action (\ref{action}). The double time integrals in the terms with coefficients $D$ and $\hat D$ in (\ref{action}) are due to the time independence of the variances (\ref{disorder-dis}).

\subsection{Details of the dynamic RG calculations}

Expressions for the different diagrams; rescaling of space, time and the fields; scaling of the model parameters. Upper critical dimensions.

The two-point correlation functions in the harmonic theory (i.e., after setting $u=0=v=D=\hat D$ in (\ref{action}) can be directly read off (\ref{action}):
\begin{eqnarray}
 \langle  \phi_i({\bf q},\omega)\hat\phi_j({-\bf q},-\omega)\rangle &=& \frac{\delta_{ij}} {-i\omega + r_0 + q^2},\\
 \langle \phi_i({\bf q},\omega) \phi_j(-{\bf q},-\omega)&=& \frac{2\Gamma\delta_{ij}}{\omega^2 + (r_0 + q^2)^2}.
\end{eqnarray}
Once the fields having support in the wavevector range $\Lambda/b$ to $\Lambda$ are integrated out, we obtain ``new'' model parameters corresponding to a modified action ${\cal S}^<$ having $\Lambda/b$ as the wavevector upper cutoff. This procedure yields Eqs.~(\ref{req}-\ref{hatDeq}). In the next step, we scale wavevectors and frequencies: $q\rightarrow bq,\; \omega \rightarrow b^z \omega$. Together with the rescaling of space(or momentum) and time(or frequency), long wavelength parts of the fields are rescaled:
\begin{eqnarray}
 \phi_i({\bf q},\omega)&=&\xi\phi_i(b{\bf q},b^z\omega),\\
 \hat\phi_i({\bf q},\omega)&=&\hat\xi\hat \phi_i(b{\bf q},b^z\omega).
\end{eqnarray}
We can now get the rescaling factors of the model parameters as follows:
\begin{eqnarray}
&& r_0'=b^2r_0^<,\\
&&\Gamma'=b^{z-2}\Gamma^<,\\
&&u'=u^<b^\epsilon,\\
&&v'=v^<b^\epsilon,\\
&&D'=D^<b^\epsilon,\\
&&\hat D= \hat D^< b^\epsilon.\label{rescale-fac}
\end{eqnarray}
To get rescaling factors in (\ref{rescale-fac}), we have demanded that the coefficients of the $\int d^dq\, d\omega \,i\omega \hat\phi_i({\bf q},\omega)\phi_i ({\bf q},\omega)$ and $\int d^dq\,d\omega \,q^2\hat\phi_i (-{\bf q},-\omega)\phi_i({\bf q},\omega)$ to unity, giving $\hat\xi = \xi b^{-z}$ and $\xi^2=b^{d+2z+2}$. Here, $\epsilon\equiv 4-d$. Thus, $d_c=4$ is the upper critical dimension of all the coupling constants $u,\,v,\,D$ and $\hat D$.

We now list the expressions for the Feynman diagrams in the main text along with their respective symmetry factors. In Fig.~\ref{vertex}, we give the graphical representations for the anharmonic vertices in (\ref{rep-free1}).

\begin{figure}[htb]
 \includegraphics[width=9cm]{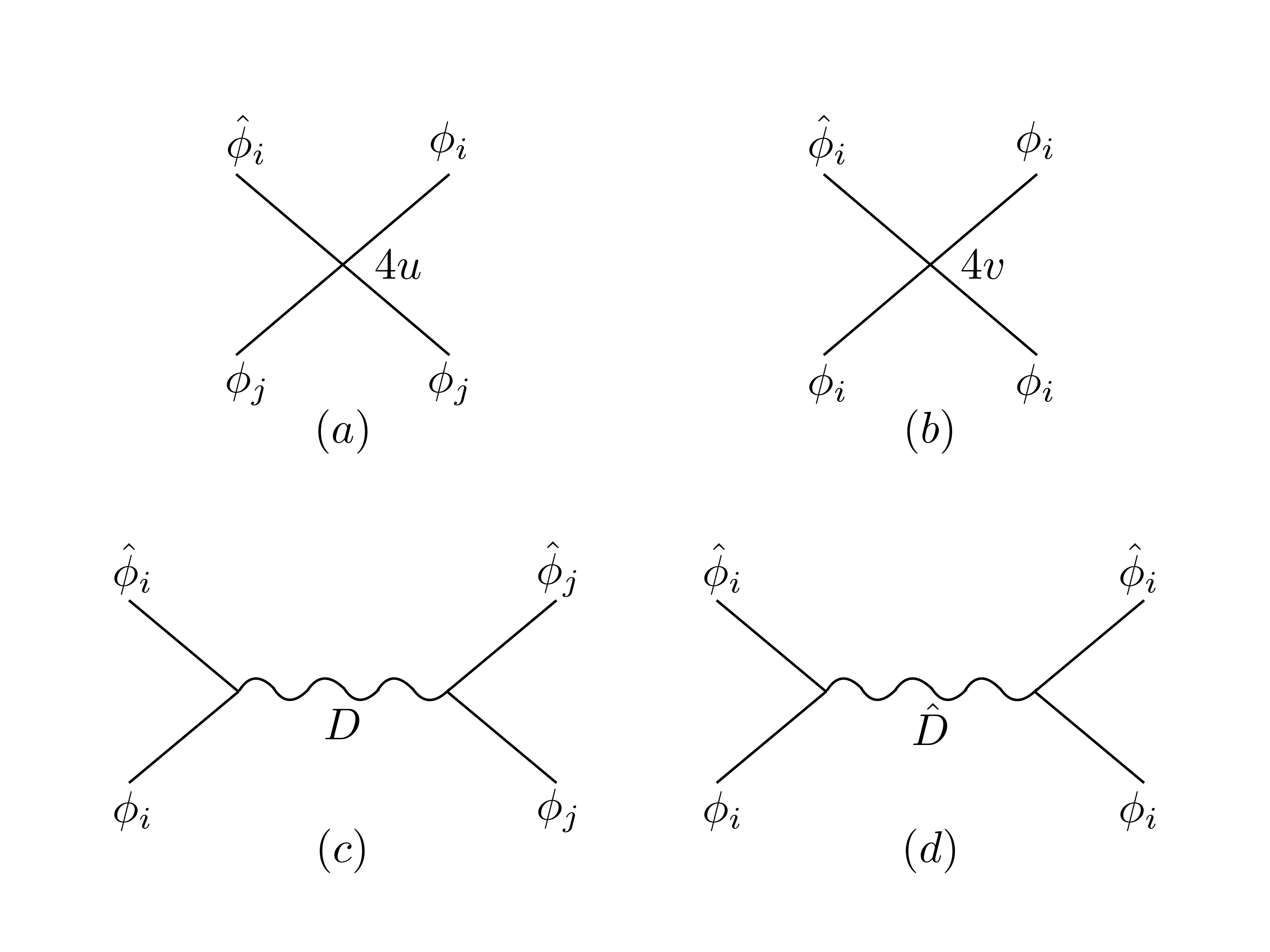}
 \caption{Vertices for the Feynman diagrams: (a) $4u \hat \phi_i\phi_j^2,\,$ (b) $4v\hat\phi_i \phi_i^3,\,$ (c)$\hat\phi_i\phi_i D\hat\phi_j\phi_j,\,$ (d) $\hat\phi_i \phi_i\hat D\hat\phi_i\hat\phi_i$.}\label{vertex}
\end{figure}

Next we give the one-loop graphical corrections to the propagator in Fig.~\ref{prop} below.


\begin{figure}[htb]
\includegraphics[width=10.0cm]{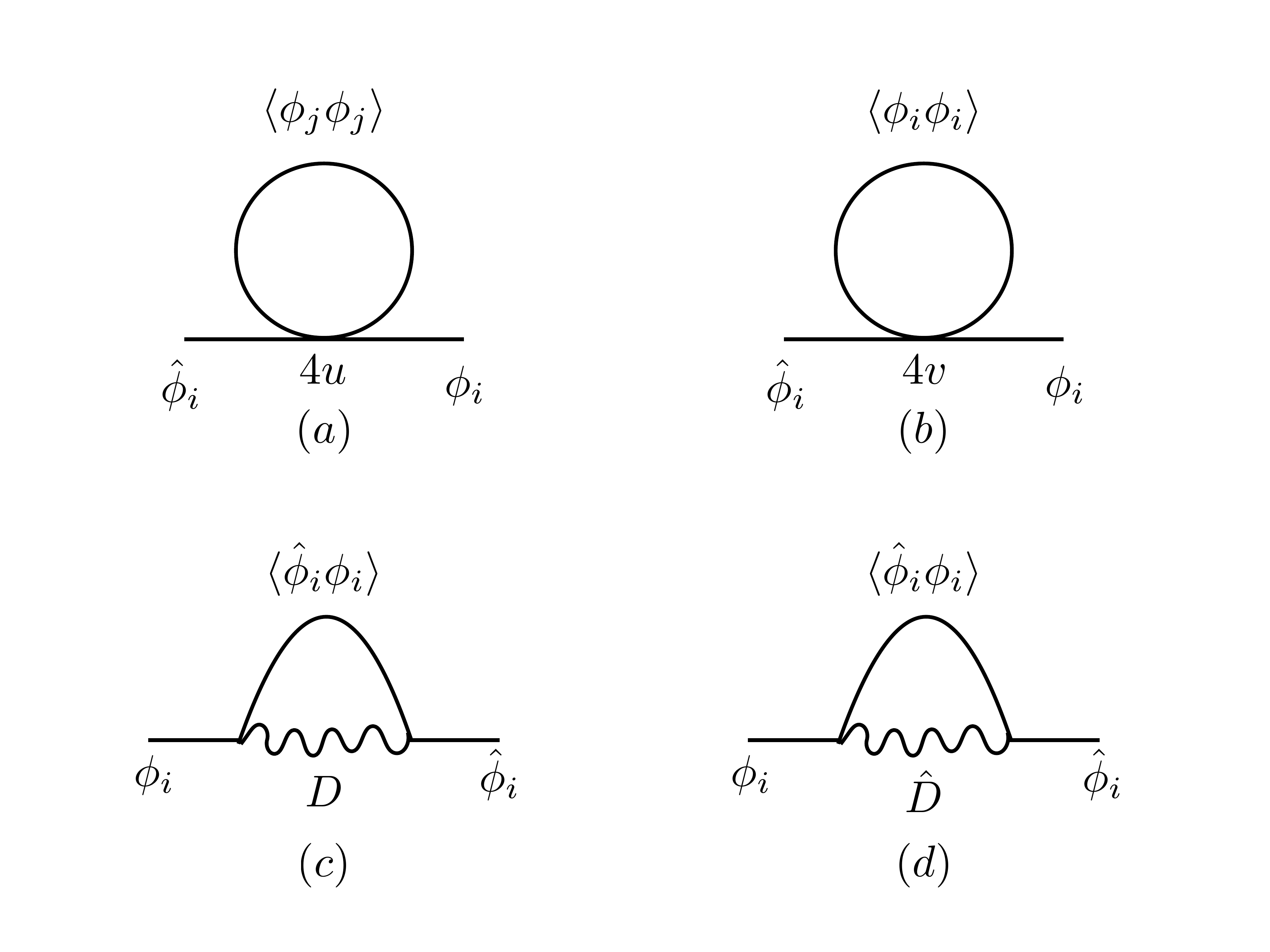}
 \caption{One-loop Feynman diagrams for the propagator.}\label{prop}
\end{figure}

We note that the diagrams (a) and (b) exist in the pure model, where as diagrams (c) and (d) appear due to the disorder. Evaluating these diagrams allow us to calculate the one-loop fluctuation-corrections to $r_0$ and $\Gamma$.

We now give below the one-loop diagrams for $u$ in Fig.~\ref{u-diag}.

\begin{widetext}
 
 \begin{figure}[htb]
 \includegraphics[width=8.9cm]{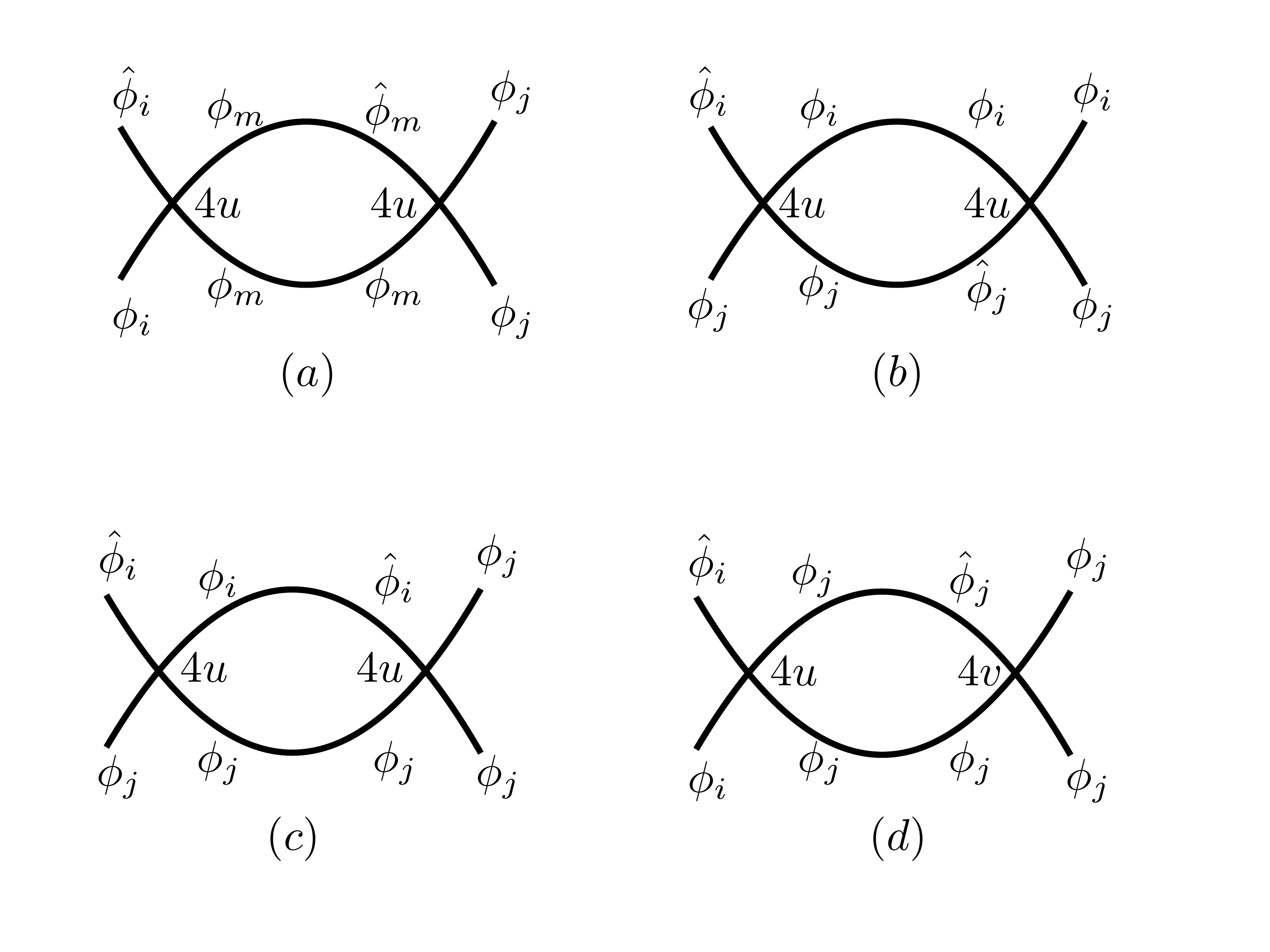}\hfill \includegraphics[width=8.9cm]{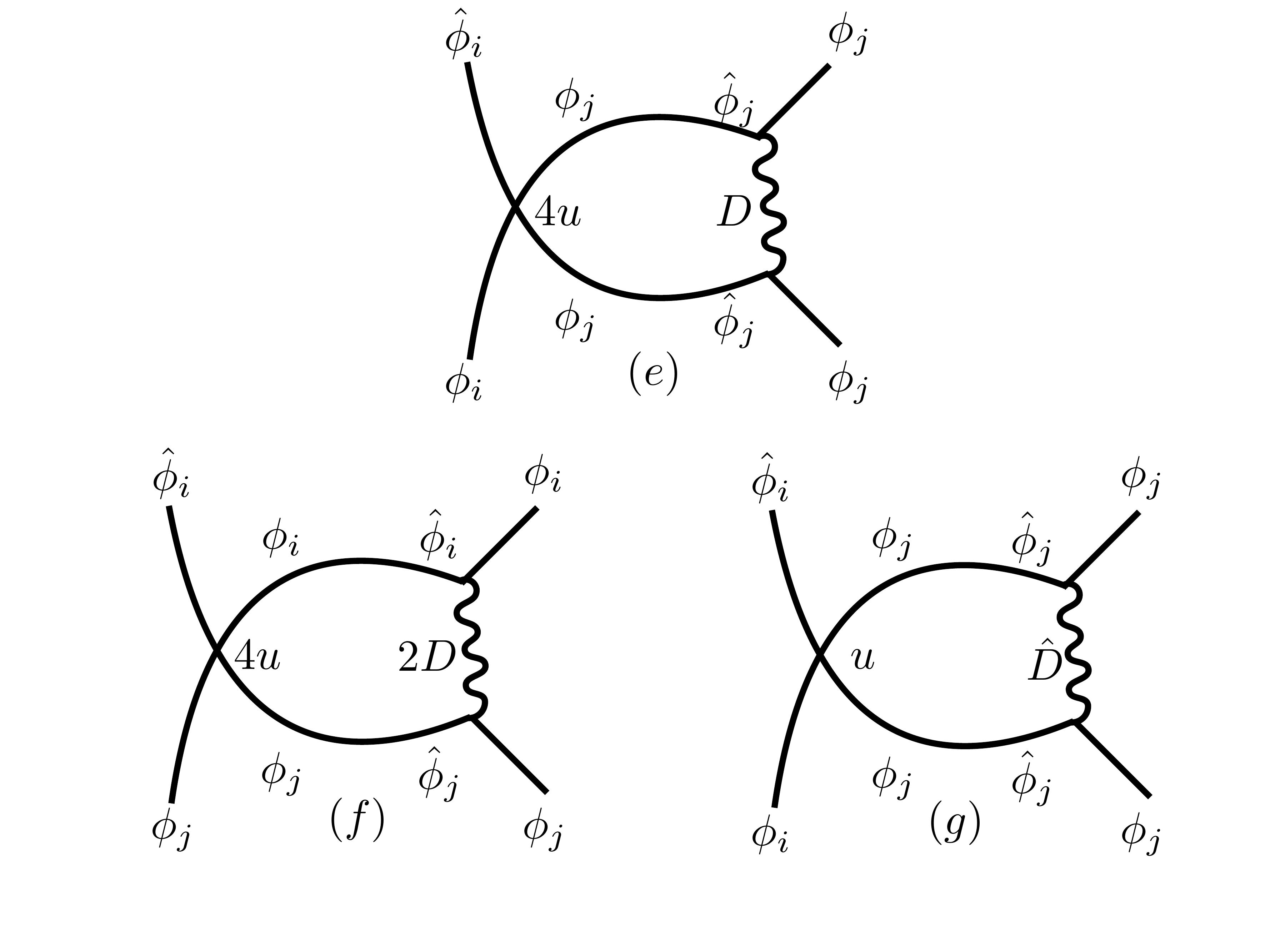}
 \caption{One-loop Feynman diagrams giving fluctuation corrections for $u$.}\label{u-diag}
\end{figure}
 
\end{widetext}

The values of the one-loop diagrams in Fig.~\ref{u-diag} are listed below.

\begin{itemize}
 \item Diagrams (a)+(b)+(c)=$-u^2[36+4(N-1)]\int\frac{d^dq}{(2\pi)}\frac{1}{(r_0+q^2)^2}.$

\item Diagram (d) = $-24 uv \int\frac{d^dq}{(2\pi)}\frac{1}{(r_0+q^2)^2}.$

\item Diagrams (e) + (f) = $12 Du \int\frac{d^dq}{(2\pi)}\frac{1}{(r_0+q^2)^2}.$

\item Diagram (g) = $4\hat D u \int\frac{d^dq}{(2\pi)}\frac{1}{(r_0+q^2)^2}.$

\end{itemize}

We now give below the one-loop diagrams for $v$ in Fig.~\ref{v-diag}.

\begin{widetext}
 
 \begin{figure}[htb]
\includegraphics[width=10cm]{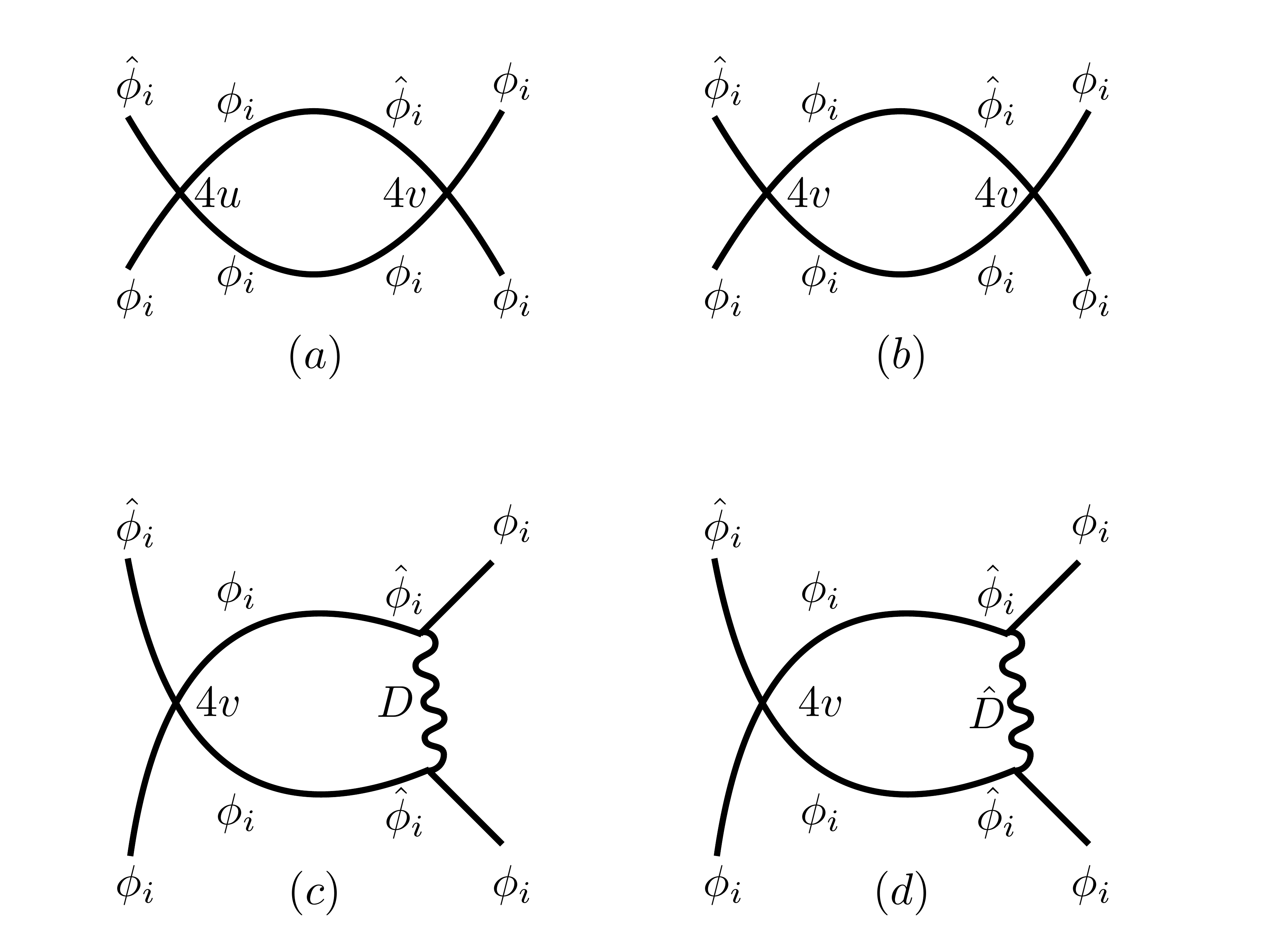}
 \caption{One-loop Feynman diagrams giving fluctuation corrections for $v$.}\label{v-diag}
\end{figure}

\end{widetext}

The values of the one-loop diagrams in Fig.~\ref{v-diag} are listed below.

\begin{itemize}
\item Diagram (a) = $-48 uv \int\frac{d^dq}{(2\pi)}\frac{1}{(r_0+q^2)^2}.$

\item Diagram (b) = $-36 v^2 \int\frac{d^dq}{(2\pi)}\frac{1}{(r_0+q^2)^2}.$

\item Diagram (c) = $ 12 v D\int\frac{d^dq}{(2\pi)}\frac{1}{(r_0+q^2)^2}.$

\item Diagram (d) = $ 12 v\hat D \int\frac{d^dq}{(2\pi)}\frac{1}{(r_0+q^2)^2}.$

\end{itemize}

\begin{widetext} 
 
We now give below the one-loop diagrams for $D$ in Fig.~\ref{D-diag}.

\begin{figure}[htb]
 \includegraphics[width=8cm]{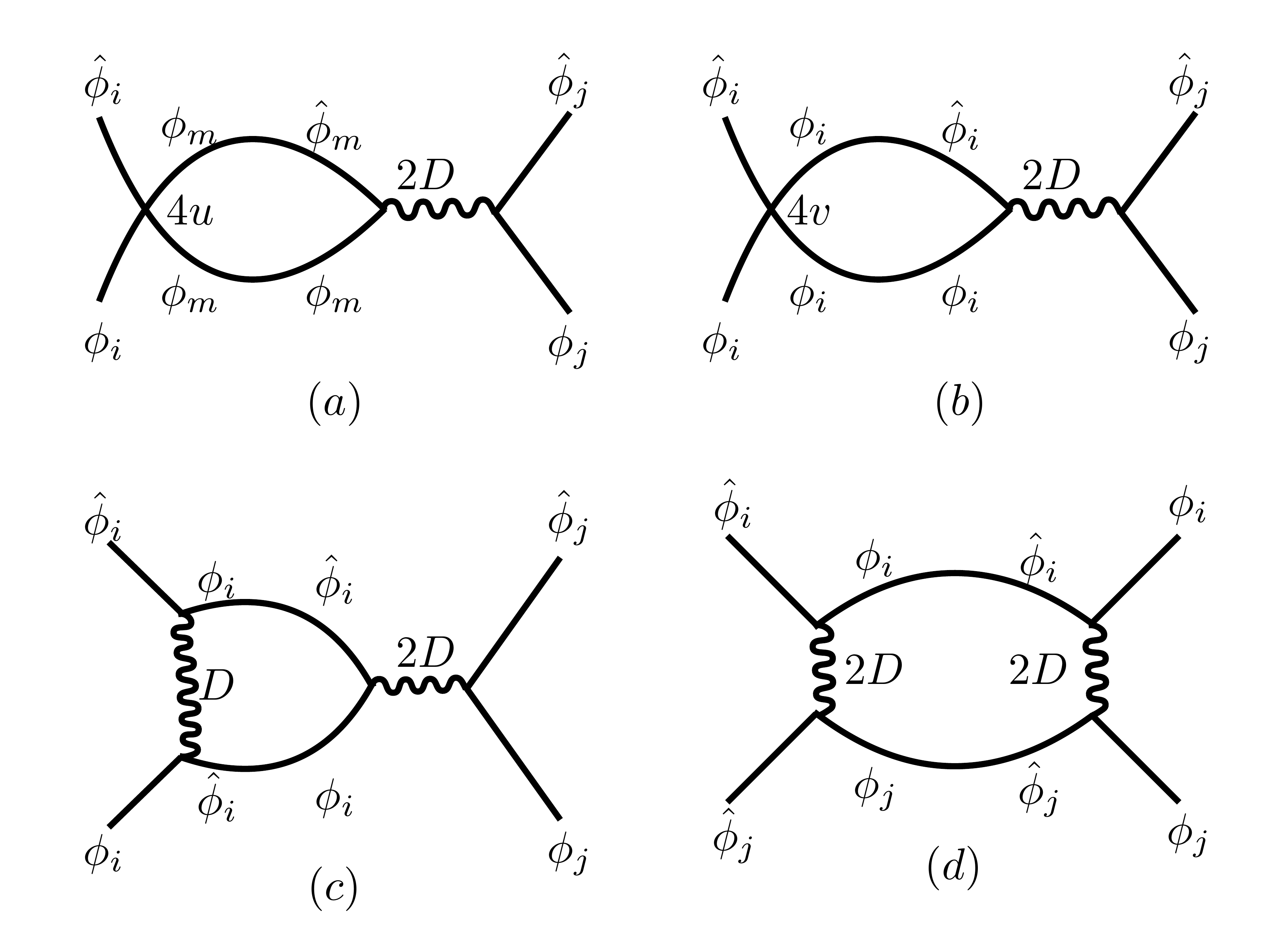}\hfill \includegraphics[width=8cm]{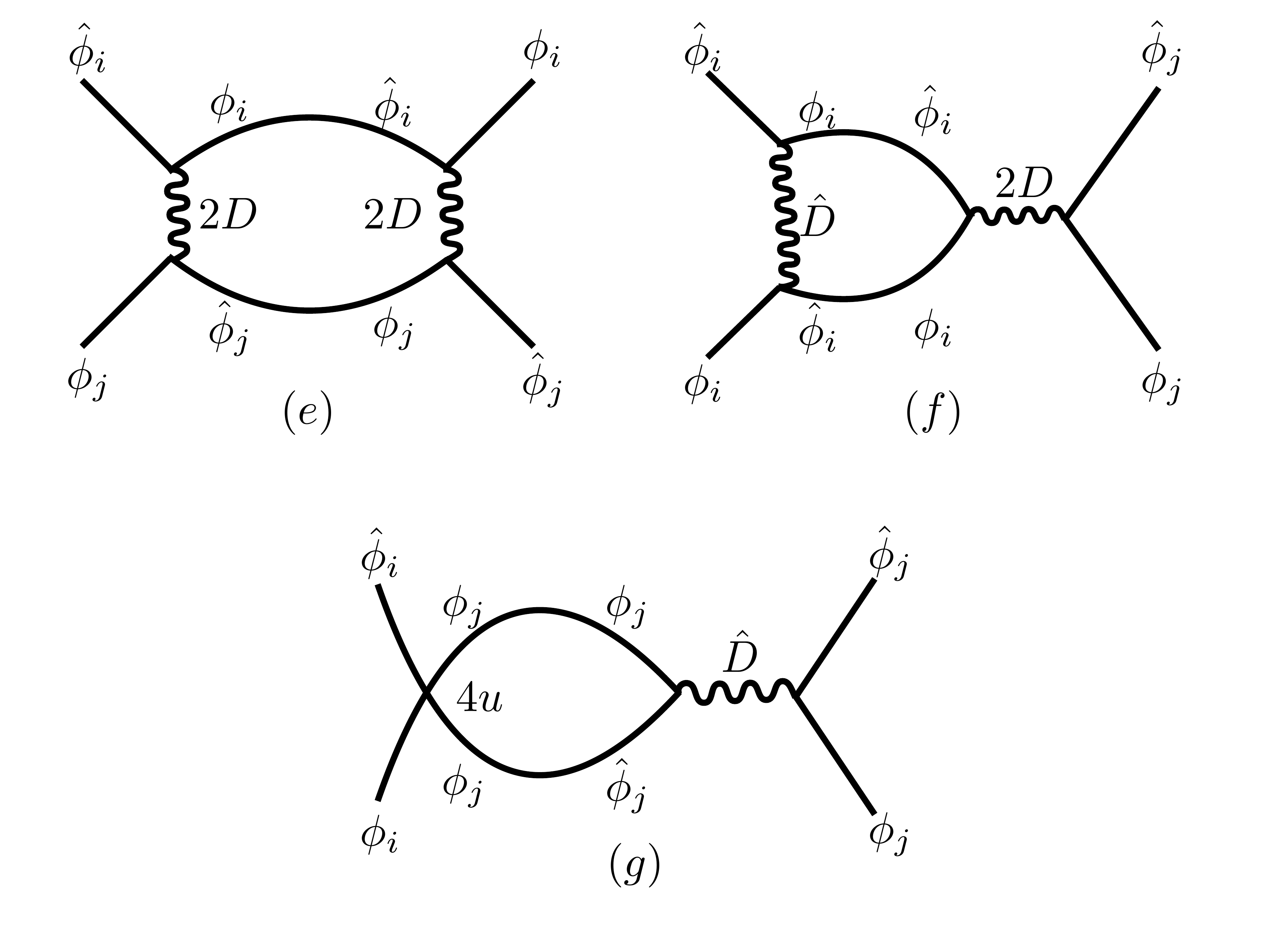}
 \caption{One-loop Feynman diagrams giving fluctuation corrections for $D$.}\label{D-diag}
\end{figure}
\end{widetext}

The values of the one-loop diagrams in Fig.~\ref{D-diag} are listed below.

\begin{itemize}
 
\item Diagram (a)=$-8(N+2)u\int\frac{d^dq}{(2\pi)}\frac{1}{(r_0+q^2)^2}.$

\item Diagram (b) = $-24 v \int\frac{d^dq}{(2\pi)}\frac{1}{(r_0+q^2)^2}.$

\item Diagrams (c) + (d) + (e) = $8D^2\int\frac{d^dq}{(2\pi)}\frac{1}{(r_0+q^2)^2}.$

\item Diagram (f) = $ 4D\hat D \int\frac{d^dq}{(2\pi)}\frac{1}{(r_0+q^2)^2}.$

\item Diagram (g) = $-8u\hat D \int\frac{d^dq}{(2\pi)}\frac{1}{(r_0+q^2)^2}.$

\end{itemize}

We now give below the one-loop diagrams for $\hat-D$ in Fig.~\ref{hat-D-diag}.

\begin{widetext}

\begin{figure}[htb]
 \includegraphics[width=8.7cm]{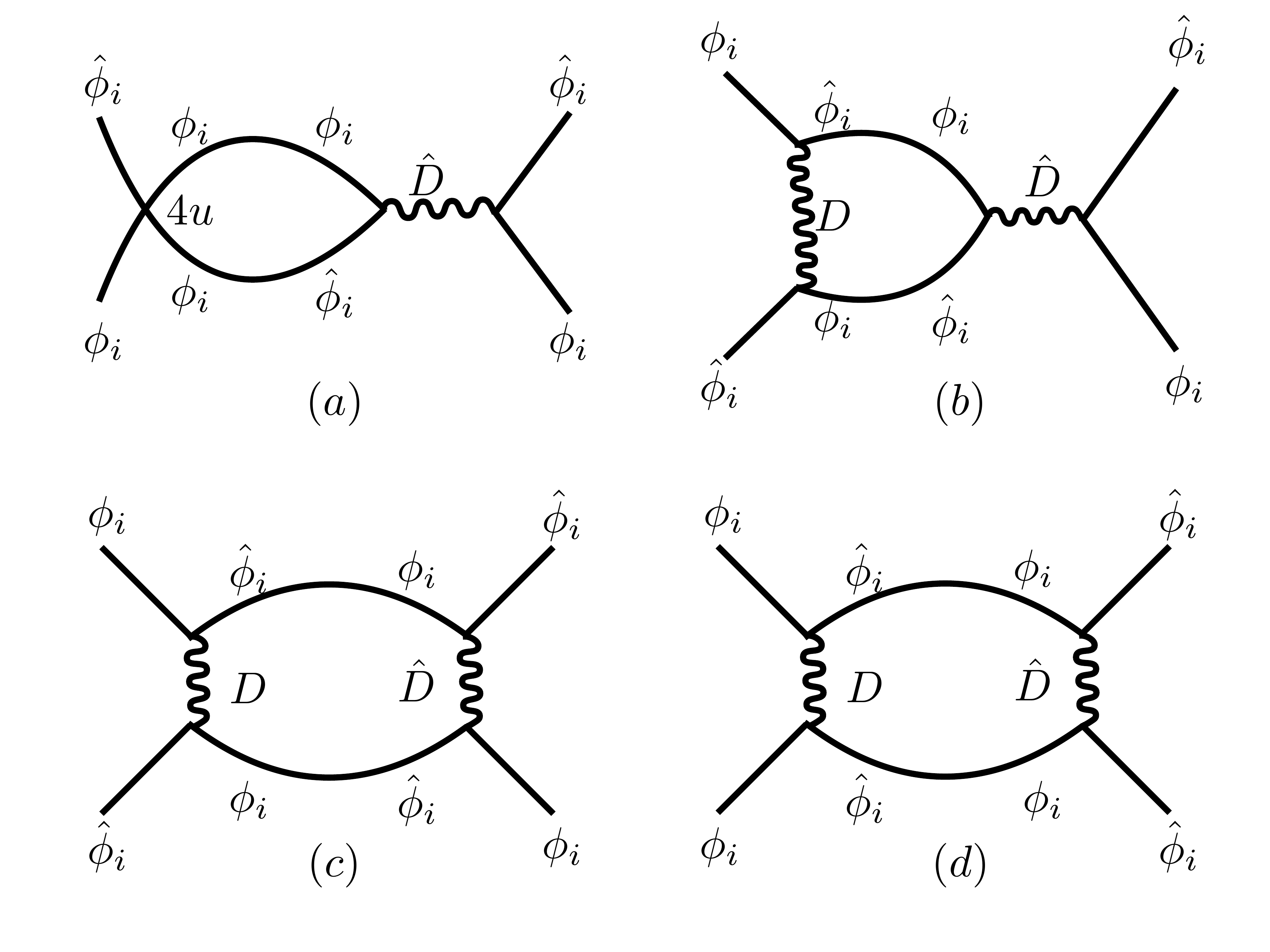}\hfill \includegraphics[width=8.7cm]{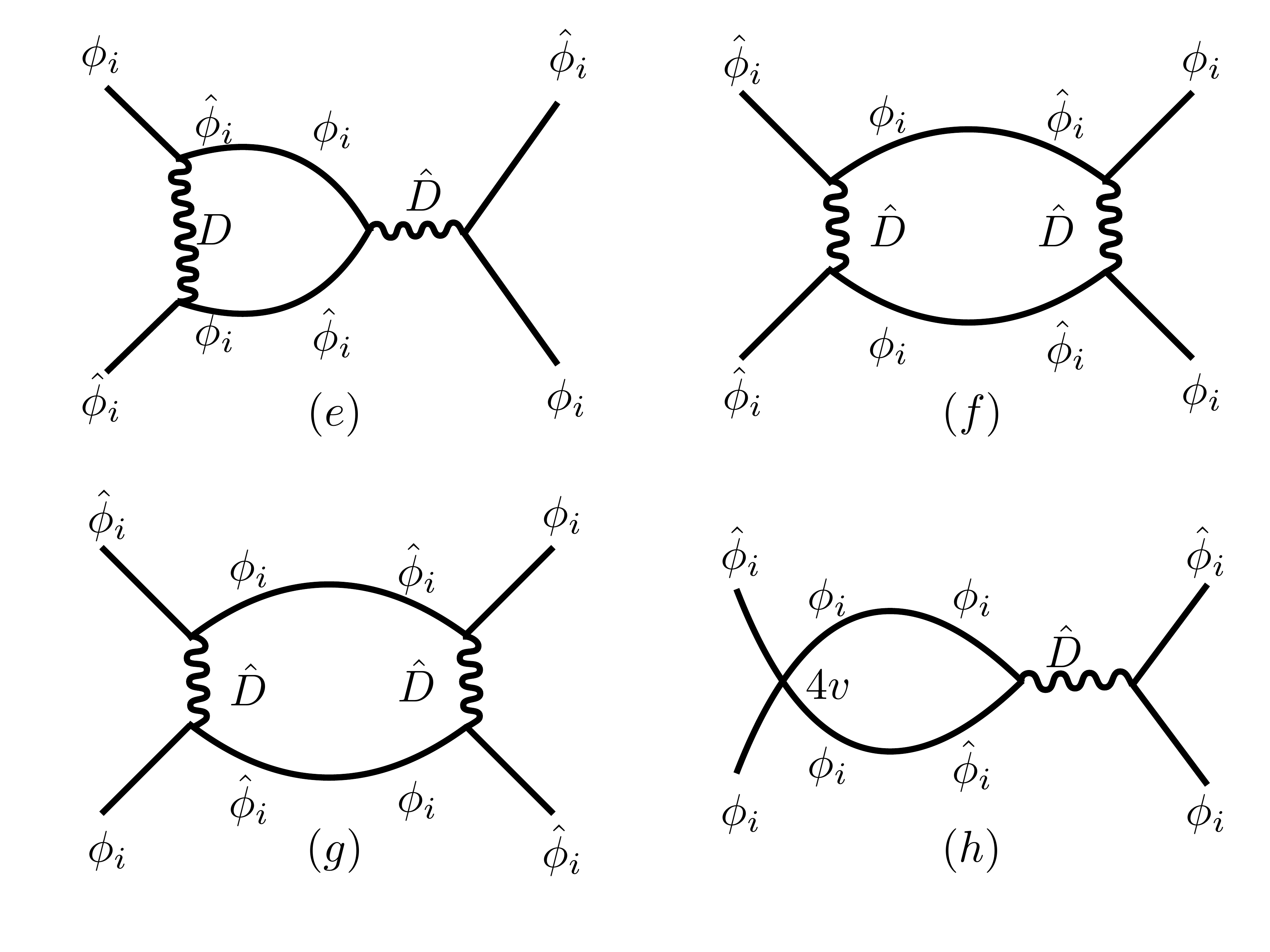}
 \caption{One-loop Feynman diagrams giving fluctuation corrections for $\hat D$.}\label{hat-D-diag}
\end{figure}

\end{widetext} 

The values of the one-loop diagrams in Fig.~\ref{hat-D-diag} are listed below.

\begin{itemize}

\item Diagram (a) = $-16 u\hat D \int\frac{d^dq}{(2\pi)}\frac{1}{(r_0+q^2)^2}.$

\item Diagrams (b) + (c) + (d)= $12 D\hat D \int\frac{d^dq}{(2\pi)}\frac{1}{(r_0+q^2)^2}.$

\item Diagrams (e) + (f) + (g) = $8\hat D^2 \int\frac{d^dq}{(2\pi)}\frac{1}{(r_0+q^2)^2}.$

\item Diagram (h) = $-24 v \hat D\int\frac{d^dq}{(2\pi)}\frac{1}{(r_0+q^2)^2}.$

\end{itemize}

\section{Calculation of the static critical exponents from the partition function}

 For the sake of completeness and the convenience of the reader, we now outline calculation of the  static RG calculation on the partition function of the model.
 In the disorder-averaged harmonic theory, i.e., with $u=0=v=D=\hat D$, the correlation function of $\phi_{i\alpha}$ in the Fourier space reads
 \begin{equation}
  \langle \phi_{i\alpha}({\bf k})\phi_{j\beta}({\bf -k})=\frac{T\delta_{ij}\delta_{\alpha\beta}} {r_0 + k^2}.
 \end{equation}

In the presence of $u,v,D,\hat D$, the correlation function of $\phi_{i\alpha}$ cannot be calculated exactly. Further, na\"ive perturbation theory produces diverging corrections to the results from the harmonic theory at the critical point. This difficulty can be circumvent by using standard perturbative RG calculations. The detailed calculation with $\hat D=0$ is available in Ref.~\cite{disorder-free}. We provide below the additional one-loop graphical corrections for $r_0,u,v,D,\hat D$ from a non-zero $\hat D$ in Fig.~\ref{r-static-diag}, Fig.~\ref{u-static-diag}, Fig.~\ref{v-static-diag}, Fig.~\ref{v-static-diag}, Fig.~\ref{D-static-diag} and Fig.~\ref{hatD-static-diag} below. The resulting flow equations are identical to those obtained above. This establishes the equivalence and direct correspondence of the dynamic perturbation theory with the static one at the one-loop order.


\begin{figure}
 \includegraphics[width=9cm]{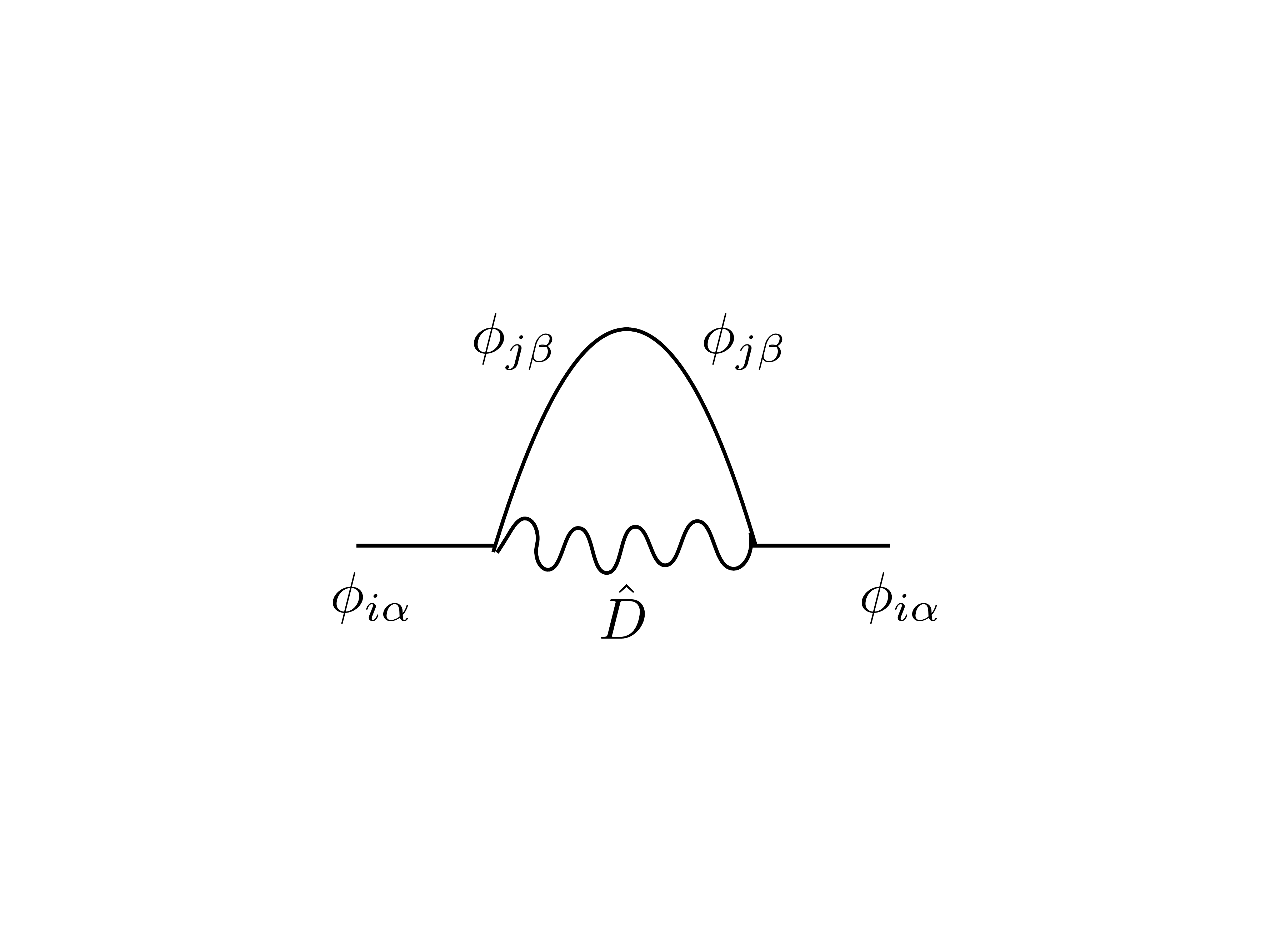}
 \caption{One-loop diagram proportional to $\hat D$ correcting $r_0$ in the static RG on the disorder-averaged free energy.}\label{r-static-diag}
\end{figure}

\begin{figure}
 \includegraphics[width=9cm]{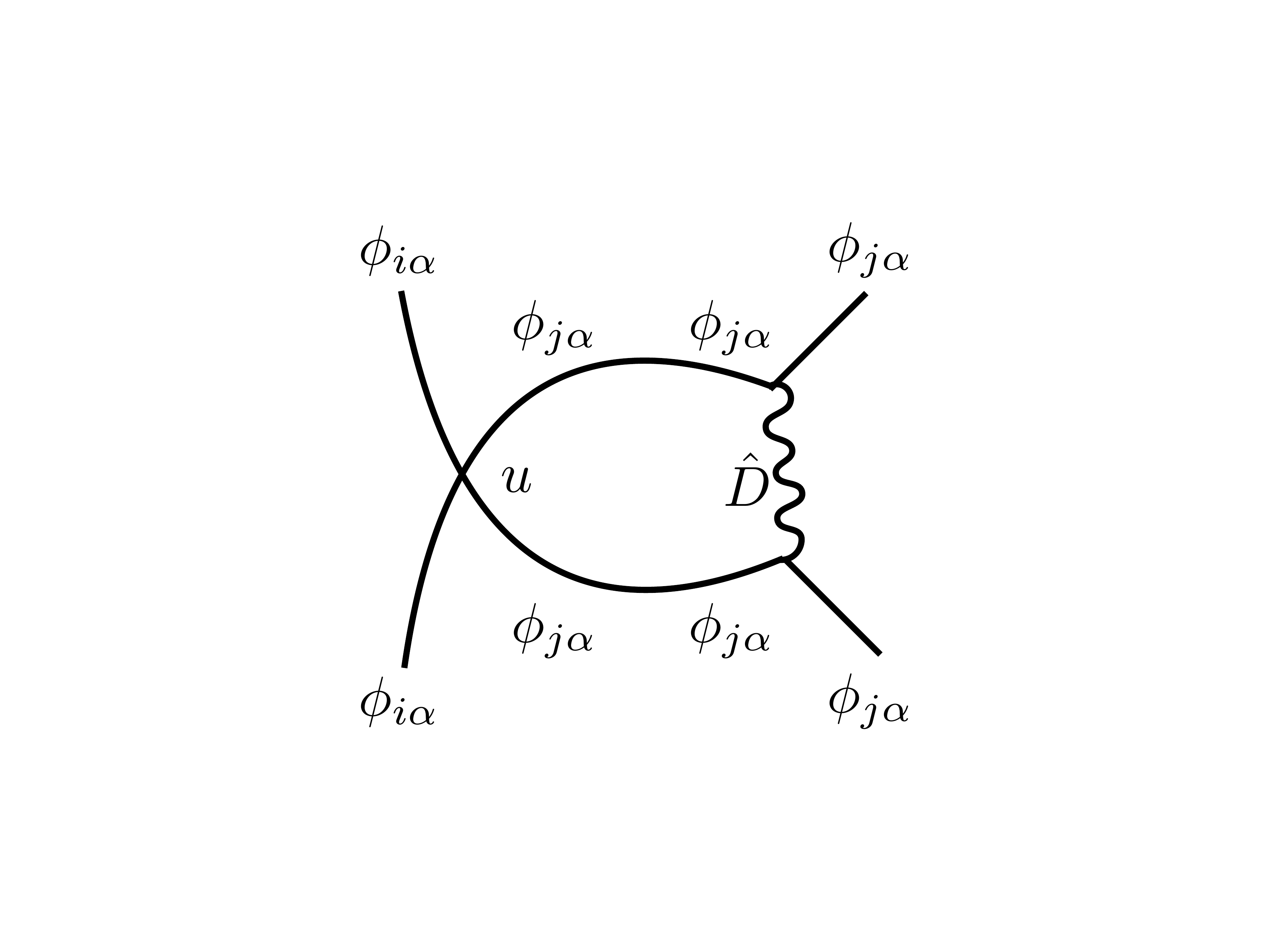}
 \caption{One-loop diagram proportional to $\hat D$ correcting $u$ in the static RG on the disorder-averaged free energy.}\label{u-static-diag}
 \end{figure}
 
 \begin{figure}
 \includegraphics[width=9cm]{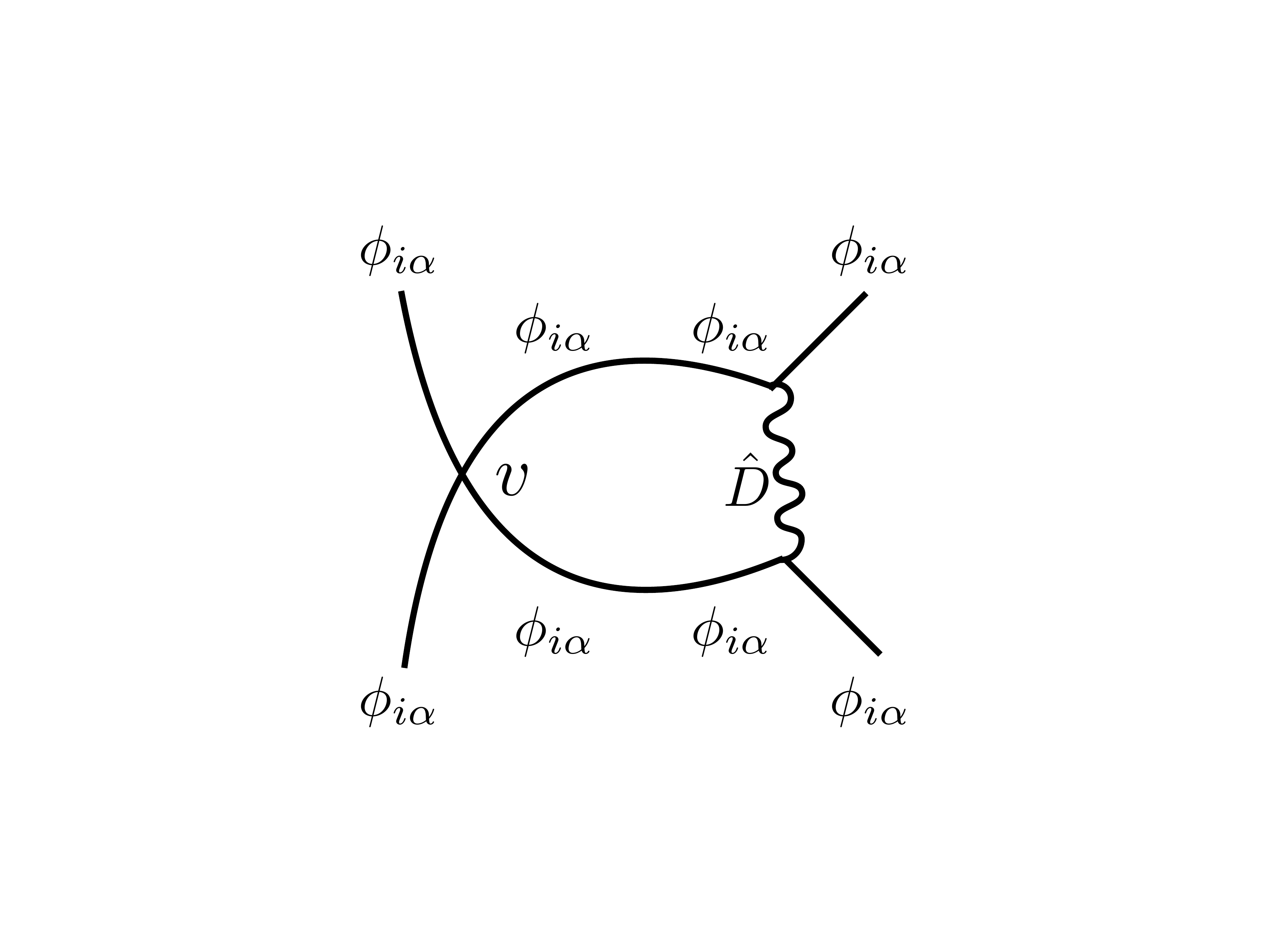}
 \caption{One-loop diagram proportional to $\hat D$ correcting $v$ in the static RG on the disorder-averaged free energy.}\label{v-static-diag}
\end{figure}

\begin{widetext}

\begin{figure}
 \includegraphics[width=12cm]{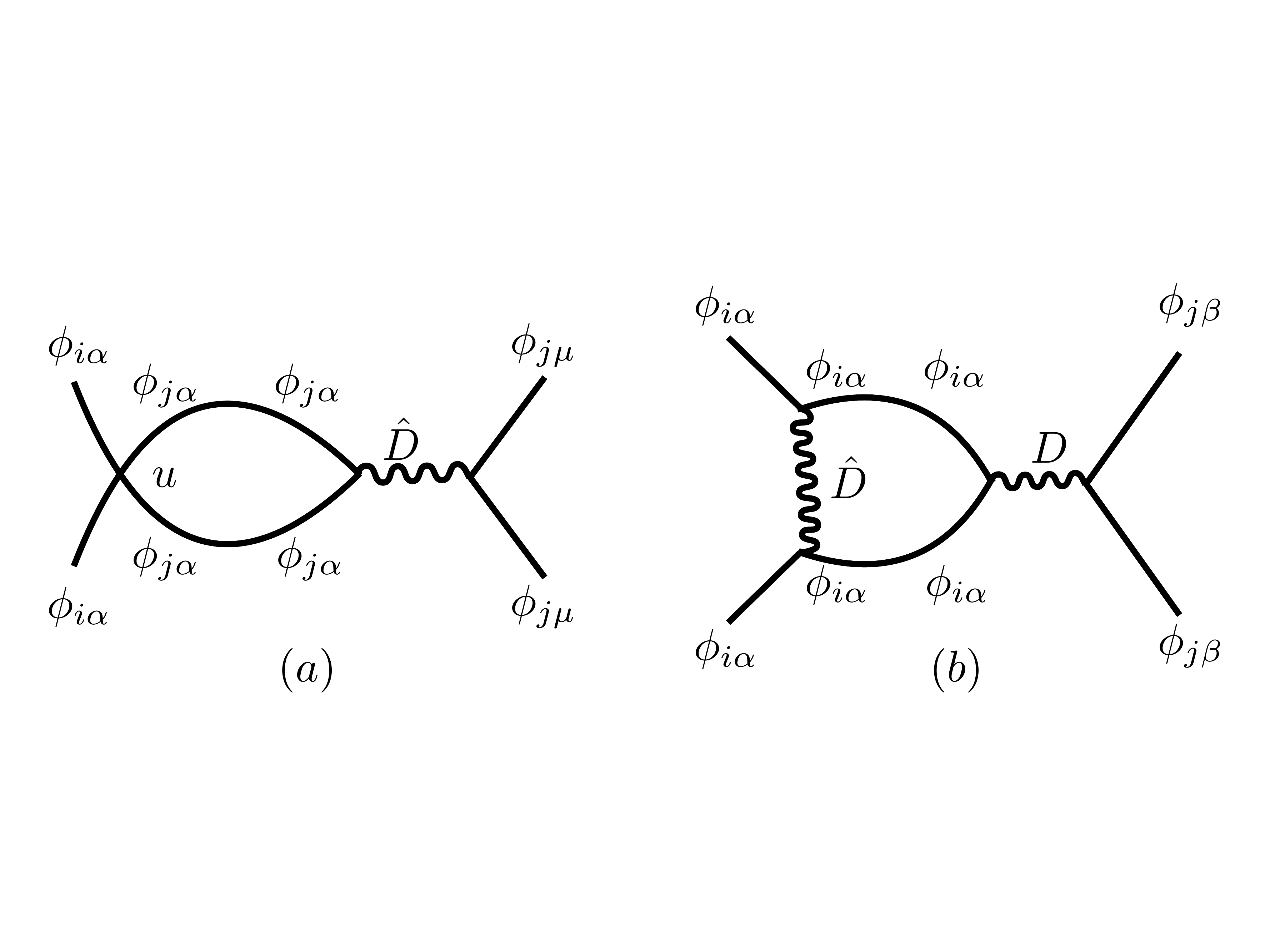}
 \caption{One-loop diagrams proportional to $\hat D$ correcting $D$ in the static RG on the disorder-averaged free energy.}\label{D-static-diag}
\end{figure}

\begin{figure}
 \includegraphics[width=8cm]{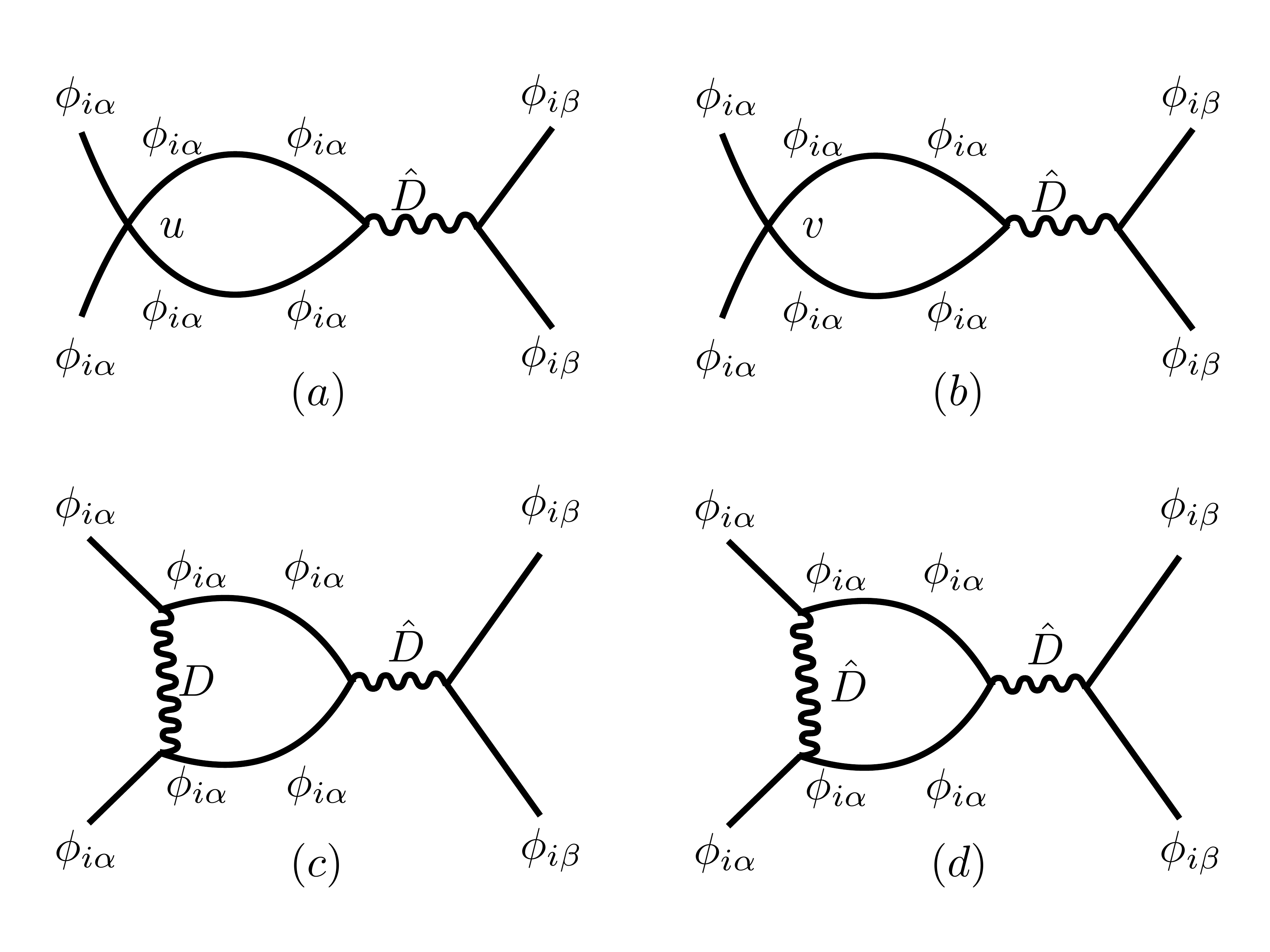}\hfill \includegraphics[width=8cm]{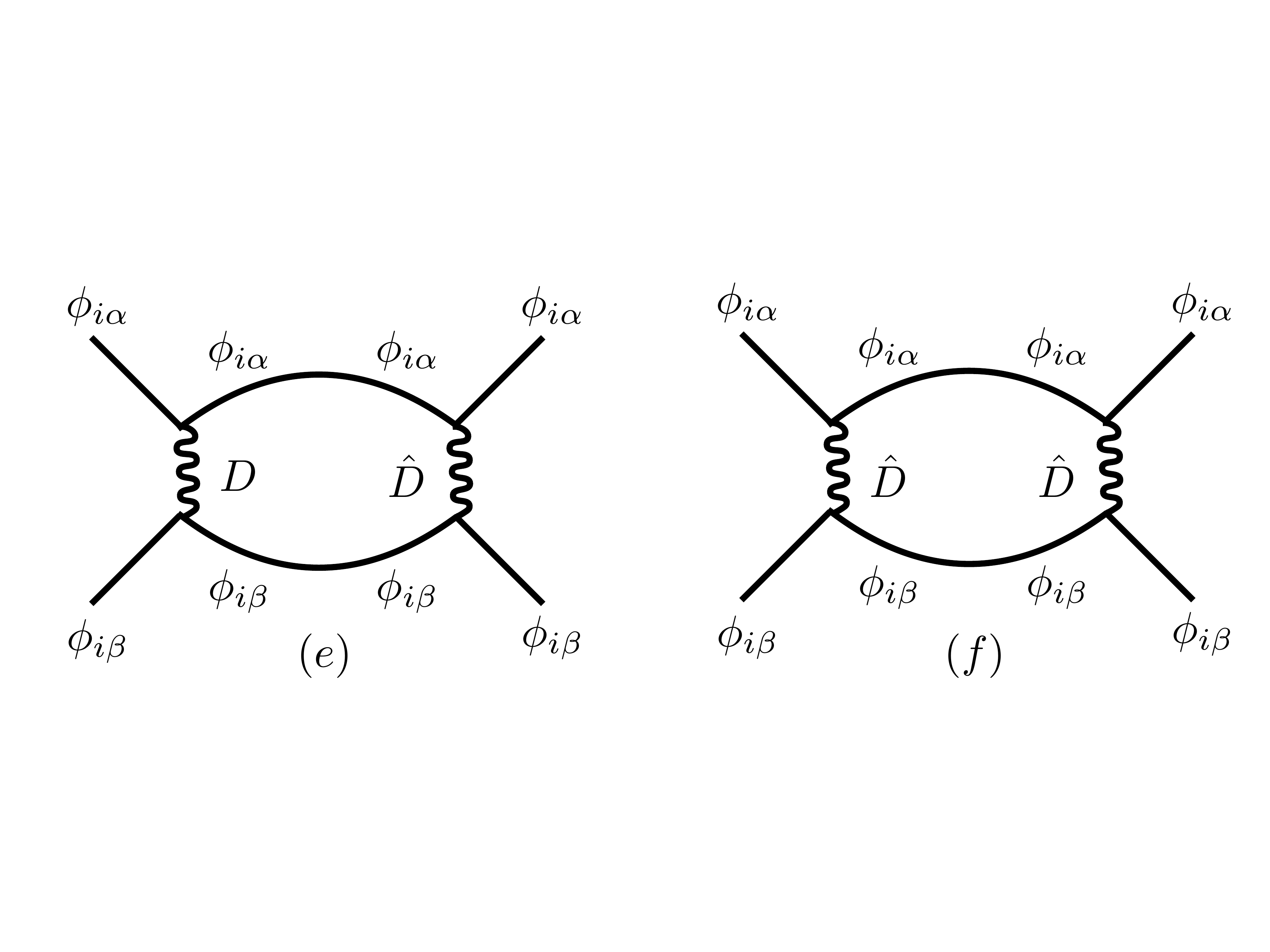}
 \caption{One-loop diagrams proportional to $\hat D$ correcting $\hat D$ in the static RG on the disorder-averaged free energy.}\label{hatD-static-diag}
\end{figure}

 \end{widetext}

\newpage



\end{document}